\documentclass[sn-mathphys,Numbered]{sn-jnl}

\usepackage{graphicx}%
\usepackage{multirow}%
\usepackage{amsmath,amssymb,amsfonts}%
\usepackage{amsthm}%
\usepackage{mathrsfs}%
\usepackage[title]{appendix}%
\usepackage{xcolor}%
\usepackage{textcomp}%
\usepackage{manyfoot}%
\usepackage{booktabs}%
\usepackage{algorithm}%
\usepackage{algorithmicx}%
\usepackage{algpseudocode}%
\usepackage{listings}%

\newcommand{\y}{y}
\newcommand{\s}{s}
\newcommand{\f}{f}

\newtheorem{theorem}{Theorem}
\newtheorem{proposition}[theorem]{Proposition}%

\newcommand{\CO}{\mbox{CO}}
\newcommand{\HCO}{\mbox{HCO}}
\newcommand{\OH}{\mbox{OH}}
\newcommand{\Hh}{\mbox{H}}
\newcommand{\HtoO}{\mbox{H$_2$O}}

\newcommand{\Cl}{\mbox{Cl}}

\newcommand{\COd}{\mbox{CO$_2$}}

\providecommand{\U}[1]{\protect\rule{.1in}{.1in}}
\providecommand{\U}[1]{\protect\rule{.1in}{.1in}}

\everymath{\displaystyle}

\usepackage{graphics,color}

\begin{document}

\title{Modeling low saline carbonated water flooding \\including surface complexes}

\author*[1]{\fnm{A.C.} \sur{Alvarez}}\email{amaury@ic.ufrj.br}
\equalcont{These authors contributed equally to this work.}

\author[2]{\fnm{J.} \sur{Bruining}}\email{J.Bruining@tudelft.nl}
\equalcont{These authors contributed equally to this work.}

\author[3]{\fnm{D.} \sur{Marchesin}}\email{marchesi@impa.br}
\equalcont{These authors contributed equally to this work.}

\affil*[1]{\orgdiv{Instituto de Computação}, \orgname{Universidade Federal do Rio de Janeiro}, \orgaddress{\street{Avenida Athos da Silveira Ramos, 274}, \city{Rio de Janeiro}, \postcode{21941-590}, \state{Rio de Janeiro}, \country{Brasil}}}

\affil[2]{\orgdiv{Lab. Fluid Dynamics}, \orgname{IMPA}, \orgaddress{\street{Estrada Dona Castorina, 110}, \city{Rio de Janeiro}, \postcode{22460-320}, \state{Rio de Janeiro}, \country{Brasil}}}

\affil[3]{\orgdiv{ Civil Engineering and Geosciences}, \orgname{TU Delft}, \orgaddress{\street{ Stevinweg 1}, \city{Delft}, \postcode{2628 CE}, \state{Delft}, \country{The Netherlands}}}

\abstract{Carbonated water flooding (CWI) increases oil production due to favorable dissolution effects and viscosity reduction. Accurate modeling of CWI performance requires a simulator with the ability to capture the true physics of such process. In this study, compositional modeling  coupled with surface complexation modeling (SCM) are done, allowing a unified study of the influence in oil recovery of reduction of salt concentration in water. The compositional model consists of the conservation equations of total carbon, hydrogen, oxygen, chloride and decane. The coefficients of such equations are obtained from the equilibrium partition of  chemical species that are soluble both in oleic and the aqueous phases. SCM is done by using the PHREEQC program, which determines concentration of the master species. Estimation of the wettability as a function of the Total Bound Product (TBP) that takes into account the concentration of the complexes in the aqueous, oleic phases and in the rock walls is performed. We solve analytically and  numerically these equations in $1-$D in order to elucidate the effects of the injection of low salinity carbonated water into a reservoir containing oil equilibrated with high salinity carbonated water.}


\maketitle
\section{Introduction}

Carbonated Water Injection (CWI) is an oil recovery technique that increases the oil production due to favorable dissolution effects causing viscosity reduction \cite{dong2011experimental,ayirala2015state}. When this method is combined with low salinity brine injection, the brine behaves like a natural solvent that enhances oil recovery \cite{christensen1961carbonated,nowrouzi2018effects}.  
In this way CWI takes advantage of brine flooding to improve oil recovery from 5 to 20\% of the oil initially in place (OIIP) \cite{foroozesh2016mathematical,kumar2017comprehensive,sheng2013enhanced}.
This happens because geochemical reactions between the injected carbonated brine and rock can alter the petrophysical properties of the reservoir.

While there are several studies showing the relevance of CWI coreflood 
for enhanced oil recovery, understanding such process is still a challenge.  Among such processes are the effects of wettability, the flow of mineral salts in water and flow of $\COd$ in the aqueous and oleic phases.  In this paper, we use multiphase compositional  modeling to quantify the oil and water saturations as well as the $\COd$ transfer between the aqueous and oleic phases. As novelty we take into account the formation of surface complexes, which participate in the mechanism of wettability modification due to charge transfer  \cite{jerauld2008modeling,appelo2005geochemistry}. With this new effect we can quantify the intrinsic nonlinear relation leading to optimal recovery conditions. We also evaluate the modifications of $pH$, of the velocity of the saline front and of the magnitude of the saturation shock. These changes are the main factors for the increase in oil productivity. 

Modeling enhanced oil recovery (EOR) by conventional fractional$-$flow theory
\cite{buc,buc1} has had numerous extensions to take into account different mechanisms that contribute to injectivity.   Some model extensions have included $\COd$  \cite{welge61,nevers64,pope81} and molecular diffusion \cite{gro87}.
Thus compositional modeling has been developed to include master chemical species in the aqueous and oleic phases \cite{sohrabi2009mechanisms,farajzadeh2012detailed}.
More recently some authors have included in the model the transport of ions \cite{alvarez2018analytical1,alvarez2018analytical,sanaei2019mechanistic}. Some authors emphasize that surface complex formation is a mechanism responsible for increased oil recovery \cite{al2015geochemical}. It has often been stated that the injection of water with low salinity and $\COd$ produces chemical reactions that change wettability of rock walls favorable to enhanced oil recovery \cite{jerauld2008modeling1}. 

Some research underlines that wettability alteration changes the oil recovery when carbonated water at low salinity is injected, see, e.g., \cite{yousef2012improved,yousef2011smart}.
Understanding this recovery method is one of the purposes of this work, for conditions typical of the Brazilian pre-salt basin.  

In \cite{hirasaki2003surface} it is highlighted that the formation of surface complexes contributes to changes in wettability. This happens because charges on the calcite and oleic surfaces have opposite signs, which contributes to squeezing out the water film; this effect forms an oil-wet surface \cite{hassan2020study}. This effect can be explained in terms of changes in $pH$ values and of larger shock amplitudes in the saturation profile. Understanding this phenomenon is an expected result of this work.

To incorporate this process, we use  multiphase compositional modeling with master species. We consider the concentrations of ions generated by all equilibrium reactions present in the carbonated brine and $\COd$ in the aqueous and oleic phases. To do so, we use a one dimensional incompressible flow model that describes two-phase flow with geochemical modeling (\cite{farajzadeh2012detailed}). These processes are studied by means of a system of balance laws for the transported quantities. 
Using this model we study the flow of oil, water and dissolved carbon dioxide in a sandstone rock.  

The geochemical data are obtained by utilizing the modeling capabilities of the program PHREEQC (acronym of pH-REdox-Equilibrium C-program). This is a computer program written in
C++  designed to perform a wide
variety of aqueous geochemical calculations (see details about its implementation in \cite{parkhurst2013user,parkhurst1999user,appelo1994cation,appelo2005geochemistry}). Furthermore, we use the program for calculation of the surface complexation in calcite rock \cite{stumm2012aquatic}. The program  determines the concentration of ions
and molecules dissolved in water inclusive carbon dioxide ($\COd$), which
is the only compound that exists in both phases. By assuming chemical equilibrium, we define the behavior of all dissolved compounds by means of four transport equations of master species, which are chloride, oil, twice oxygen minus hydrogen and inorganic carbon.

We also consider the ions ($\Hh^+$, $\OH^-$, $\CO_3^{2-}$, $\HCO_3^-$, $\Cl^-$), water ($\HtoO$) and the sorbed species $oil_{s,NH^{+}}$, $oil_{w,COOH}$, $Cal_{s,OH}$, $Cal_{w,CO_{3}H}$, $oil_{s,N}$, $oil_{w,COO^{-}%
}$, $oil_{w,COOCa^{+}}$, $Cal_{s,OH_{2}^{+}}$, $Cal_{s,CO_{3}^{-}}$,
$Cal_{w,CO_{3}^{-}}$, $Cal_{w,CO_{3}Ca^{+}}$ \cite{brady2012surface}. The species with complexes lead to what we call the surface complexes$-$chloride ionic carbon dioxide$-$oil$-$water SC\--CLICDOW model. This model summarizes the most relevant concepts described in \cite{bryant1986interactions,bryant1987mineral,helfferich1989theory,lake1989enhanced,lake2014}. To illustrate these effects we inject water with the same $pH$ but with lower salinity than the initial salinity in the core.

In this study,  we use SCM through the geochemistry PHREEQC program. This solver is utilized to calculate the concentration of the complexes and to predict the wettability alteration of minerals through the Total Bound Product (TBP). Details of the  procedure employed here can found in \cite{mehdiyev2022surface}. SCM is a chemical equilibrium technique used to model the interactions of water, oil, brine and rock \cite{lutzenkirchen2006surface,elakneswaran2017surface,elakneswaran2017surface}. Such a method has been used in several studies for characterizing the surface adsorption phenomenon \cite{marmier1997surface,sanaei2019investigation} and  determining the wettability of minerals at reservoir conditions \cite{ginn2020effects,bordeaux2021improvements,mehdiyev2022surface}.

In these works the model is based on adsorption of aqueous solute into surface functional groups characterized by a set of chemical reactions. Moreover, some experi\-ments indicate that since carbonates are salt-type minerals, their surface reactivity is different from sandstone and clays. Another conclusion is that dissolution and precipi\-tation interfere on ion adsorption dynamics at functional sites.
The main effects that influence the oil-brine-rock interaction are
the brine chemistry and the oil composition, i.e., acid and base number \cite{dubey1993base}.

 Several works describe how to quantity the relationship between salt concentration and
 water flooding process, e.g., \cite{jerauld2008modeling,omekeh2012modeling,al2015novel}. In this paper we use a similar idea, which is proposed by \cite{jerauld2008modeling} utilizing the relative permeability functions based on Corey's co\-rre\-la\-tion.

Based on Gibbs rule and assuming that the sodium and chloride concentrations are a\-ppro\-ximately equal, we consider four balance laws, i.e., total carbon, hy\-dro\-gen\-$-$o\-xy\-gen, chloride and decane. Each one consists of three terms, i.e., accumulation, convection and the combination of molecular and capillary diffusion. In this way, we obtain analytically the Riemann solution, which consists basically in applying the method of characteristics (MOC) through the wave curve method. In this work we use this method to seek analytical solutions of the SC$-$CLICDOW model similar to the model treated
in \cite{nevers64,welge61,buc,buc1,pope81,dumore,johns93,alvarez2018analytical1}. The Riemann solution consists of a concatenation of spreading and shock  waves, implementing certain admissibility conditions (\cite{lax60,olei63,glimm65,Liu74,Liu75}).

A Riemann solver for the proposed geochemical model is developed to quantify the geochemical processes of 
water injection with $\COd$ and low salinity in a carbonated reservoir. We also take into account the surface complexes formation as a mechanism to change wettability.

  A Riemann solver is developed to  automate the construction of solution paths. To do so, we take into account the bifurcation structures, which are not part of the classical fractional flow method used by \cite{pope81}. We also provide comparisons with numerical solutions obtained by means
of a commercial program (COMSOL). This procedure is extremely useful because it allows to include in a unified manner the geochemistry, the equilibrium reactions and the charge balances. Also, the method  serves to study di\-ffe\-rent situations and to include more chemical species in the system. Using 
the extended Gibbs rule (see Eq. \eqref{gibbsphaserule}), we reduce the mathematical complexity associated with considering the large number of physical constraints and parameters that are included in the geochemistry program PHREEQC.

We use a similar methodology as developed in \cite{alvarez2018analytical,alvarez2018analytical1} where the extended Gibbs phase rule was used to focus on the principal chemical species and to incorporate the geochemistry of the oil recovery. This circumvents simultaneous solution of the transport equations, of the equilibrium relations and of the effect of low salt concentrations.
	
The aims on this paper are 
(1) to quantify how the presence of surface complexes affects wettability (2) to understand how changes in the salt concentration affects relative permeability and as a consequence of the injectivity (3) to analyze the wave structure of the solutions (e.g., the occurrence of a pH wave embedded in a constant pH flood), front salt formation and a jump in water saturation.  The Riemann solution confirms that pH variations occur with various numerical schemes, inclusive discontinuous Galerkin, which is expected when surface complexes are formed. 

The additional advantage of the Riemann solution is that it can be used
	to perform a bifurcation analysis, and to make an inventory of the possible
	qualitatively different solutions. The bifurcations occur at coincidence  and inflection loci. Clearly, bifurcations are essential to build the analytical solution, as well as to determine the location
	 where qualitative changes of the behavior of the solution are expected. This determination is a useful tool for  mathematical modeling in oil recovery.

By means of numerical and analytical methods we aim in this paper at quan\-ti\-fying the recovery improvement when carbonated water at low salinity is injected in a reservoir that contains carbonated brine in equilibrium with an oleic phase and carbon dioxide. 

For the fulfillment of the objectives our paper is organized as follows.
Section II gives the physical model and  the equilibrium equations considered in this study. In Section III the derivation of the mass balance equations for master species are presented. Moreover, Corey parameters of the flow functions depending on salt concentration is presented. In  Section IV a summary of the surface complexation model is presented together with values of the parameters considered in this study. Besides, re\-gre\-ssion formulas for parameters of the system of conservation laws depending on pH and chloride concentration are presented. Furthermore, a procedure for the calculation of wettability depending on TBP is described.
Section V describes the Riemann solver and the strategy to obtain the Riemann solution.  Section VI gives the results in terms of the $pH$, the chloride concentration, the water saturation and the total velocity. The calculation suggests that a low salinity carbonated water flood improves the recovery because it admits a high dissolved concentration of carbon dioxide.

\section{Physical model}

We consider the injection of low salinity brine (0.5 $mol/liter$ NaCl, saturated with $\COd$ at a $pH=4.0$) into an inert rock filled with an oleic phase. Injection and initial fluids contain carbon dioxide and other unrelated ions such as sodium chloride. We assume chemical equilibrium in both the aqueous phase and the oleic phase \cite{alvarez2018analytical1}. The ions and water are only present in the aqueous phase, decane is only present in the oleic phase. Indeed, dissolution  of oil in the aqueous phase is disregarded. The solubility of carbon dioxide decreases dramatically
at high salt concentration. We assume that the flow is governed by Darcy's law. The temperature is chosen to be 39$^{o}$C because literature data is available and the pressure is chosen to be well above the pressure at which a gaseous phase can exist.

We apply the extended Gibbs phase rule to determine the number
of degrees of freedom $n_f$. This rule states (see, e.g.,
\cite{hona86,merkel2005groundwater}) that this number is given
by
\begin{equation}
n_{f}=N_s+n_{s}-N_r-n_{c}+2-p, \label{gibbsphaserule}%
\end{equation}
where $N_{s}$ is the number of dissolved chemical species, $n_{s}$ is the
number of surface species, $N_{r}$ is the number of chemical reactions and
$n_{c}$ is the number of constraints, e.g., the charge balance. The number $2$
represents the temperature and pressure and $p$ the number of phases.

We follow Appelo and Parkhurst \cite{parkhurst2013user} and
\cite{appelo2005geochemistry} and obtain with the geochemistry program
PHREEQC, when we add water, $CaCO_{3}\left(  solid\right)  $ and $NaCl$, that
there are fifteen different chemical species $\left(  N_{s}=15\right)  $ with
molar concentrations in the aqueous phase: $c_{a,CO_{2}},$ $c_{a,CO_{3}^{2-}%
},$ $c_{a,HCO_{3}^{-}},$ $c_{a,CaHCO_{3}^{+}},$ $c_{a,CaCO_{3}},$
$c_{a,NaCO_{3}^{-}},$ $c_{a,NaHCO_{3}},c_{a,H_{2}O},$ $c_{a,H^{+}},$
$c_{a,OH^{-}},$ $c_{a,CaOH^{+}},c_{a,Ca^{2+}},c_{a,Cl^{-}},c_{a,Na^{+}}$ and
the alkane $\left(  A\right)  $ concentration in the oleic phase $c_{o,A}$ .
The alkane only occurs in the oleic phase, whereas all the other components
occur only in the aqueous phase. $CaCO_{3}$ occurs both in the solid phase with concentration
$c_{r,CaCO_{3}}$ and in the aqueous phase with concentration $c_{a,CaCO_{3}}$.
\ In addition we have $n_s=11$ sorbed species $oil_{s,NH^{+}}$,
$oil_{w,COOH}$, $Cal_{s,OH}$, $Cal_{w,CO_{3}H}$, $oil_{s,N}$, $oil_{w,COO^{-}%
}$, $oil_{w,COOCa^{+}}$, $Cal_{s,OH_{2}^{+}}$, $Cal_{s,CO_{3}^{-}}$,
$Cal_{w,CO_{3}^{-}}$, $Cal_{w,CO_{3}Ca^{+}}.$ \ There are seven surface
reactions taking into account complexes. $\ $These surface reactions are \cite{brady2012surface}
\begin{align}
oil_{s,}{}_{^{NH^{+}}}\rightleftharpoons oil_{s,N}+H^{+}  &  \quad\log
K_{-11}=-6.0\nonumber\\
oil_{w,COOH}\rightleftharpoons oil_{w,COO^{-}}+H^{+}  &  \quad\log
K_{-21}=-5.0\nonumber\\
oil_{w,COOH}+Ca^{2+}\rightleftharpoons oil_{w,COOCa^{+}}+H^{+}  &  \quad\log
K_{22}=-3.8\nonumber\\
Cal_{s,OH}+H^{+}\rightleftharpoons Cal_{s,OH_{2}^{+}}  &  \quad\log
K_{13}=11.8\nonumber\\
Cal_{s,OH}+HCO_{3}^{-}\rightleftharpoons Cal_{s,CO_{3}^{-}}+H_{2}O  &
\quad\log K_{33}=5.8\nonumber\\
Cal_{w,CO_{3}H}\rightleftharpoons Cal_{w,CO_{3}^{-}}+H^{+}  &  \quad\log
K_{-14}=-5.1\nonumber\\
Cal_{w,CO_{3}H}+Ca^{2+}\rightleftharpoons Cal_{w,CO_{3}Ca^{+}}+H^{+}  &
\quad\log K_{24}=-2.6 \label{brady}%
\end{align}
We use four different sorption sites. Each of the sorption sites can receive sorption
molecules of two types, called $weak$ and $strong$. For instance the strong adsorption sites in oil
can receive both $oil_{s,NH^{+}}$ and $oil_{s,N}$, so that the sum of the
sorbed concentration at each sorption site is a given constant. A similar
situation occurs for the weak adsorption sites on oil $\left(  oil_{s}%
\right),$ but now with three surface species. The concentration of
$CaCO_{3}$ in the solid phase is known and constant.  Thus taking into account the charge balance equation we have $n_c=5$ constraints.

We notice that the species can be derived from seven master species, i.e.,
$C\left(  4\right)  ,$ $H\left(  1\right)  ,$ $O\left(  -2\right)  ,$ $Ca(2),$
$Na(1),$ $Cl\left(  -1\right)  $, $C\left(  -4\right)  $, where the value in
parentheses denotes the constant valence of the species. We note that $C\left(
-4\right)  $ is used to denote organic carbon as opposed to $C\left(
4\right)  ,$ which denotes inorganic carbon.

We consider the following eight  equilibrium
reactions in the aqueous phase (thus we have $N_{r}=15 $)%
\[
\left(  CO_{2}\right)  _{aq}+H_{2}O\rightleftharpoons HCO_{3}^{-}+H^{+}%
\]
\[
HCO_{3}^{-}\rightleftharpoons CO_{3}^{2-}+H^{+}%
\]%
\[
H_{2}O\rightleftharpoons OH^{-}+H^{+}%
\]%
\[
\left(  CaCO_{3}\right)  _{aq}\rightleftharpoons Ca^{2+}+CO_{3}^{2-}
\label{CaCO3eq}%
\]%
\[
Ca^{2+}+H_{2}O\rightleftharpoons CaOH^{+}+H^{+}%
\]%
\[
CO_{3}^{2-}+Ca^{2+}+H^{+}\rightleftharpoons CaHCO_{3}^{+}%
\]%
\[
Na^{+}+CO_{3}^{2-}\rightleftharpoons NaCO_{3}^{-}%
\]%
\[
Na^{+}+HCO_{3}^{-}=NaHCO_{3} \label{equilibriumrelation}%
\]%

We dropped the subscript $\left(  aq\right)  $ on all compounds except
$CaCO_{3}$ and $CO_{2}$ as we assume that they only occur in the aqueous
phase. All possible equilibrium equations can be derived from these eight
equilibrium reactions.

Thermodynamic equilibrium requires that the chemical potential of $\left(
CaCO_{3}\right)_{r}$ in the solid phase is equal to the chemical potential
of $\left(  CaCO_{3}\right) _{aq}$ in the aqueous phase. \ This can be
represented by
\begin{equation}
\left(  CaCO_{3}\right)  _{r}\rightleftharpoons\left(  CaCO_{3}\right)  _{aq}.
\label{CaCO3equil}%
\end{equation}
In the same way the chemical potential of carbon dioxide in the aqueous phase
is equal to the chemical potential in the oleic phase. This can be
represented as
\begin{equation}
\left(  CO_{2}\right)  _{o}\rightleftharpoons\left(  CO_{2}\right)  _{aq}.
\label{CO2equil}%
\end{equation}
As we consider solid, aqueous and oleic phases, the number of phases $p$ is $3$. We have the charge balance equation given by

\begin{equation}
\left(
\begin{array}
[c]{c}%
2c_{a,CO_{3}}+oil_{w,COO^{-}}+c_{a,HCO_{3}}+c_{a,OH}+c_{a,NaCO3}%
+c_{a,Cl}+Cal_{s,CO_{3}^{-}}+Cal_{w,CO_{3}^{-}}\\
=2c_{a,Ca}+oil_{s,}{}_{^{NH^{+}}}+c_{a,H}+c_{a,Ca(HCO_{3})} \nonumber \\
+c_{a,Na}%
+c_{a,CaOH}+oil_{w,COOCa^{+}}+Cal_{s,OH_{2}^{+}}+Cal_{w,CO_{3}Ca^{+}}%
\end{array}
\right)  . \label{chbal}%
\end{equation}
This charge balance equation can also be derived from the mass balance
equations. Thus, we adopt the charge balance equation allowing one mass balance equation to be removed.  

Following Gibbs rule described in \eqref{gibbsphaserule} the number $n_{f}$ of degrees of freedom is
\begin{equation}
n_{f}=N_s+n_{s}-N_r-n_{c}-p+2=15+11-15-5-3+2=5. \label{Gibbs}%
\end{equation}
 In Eq. \ref{Gibbs}
we use that $N_s+n_{s}-N_r-n_{c}=6$. Given the temperature and pressure,
we only need three concentrations to specify the composition of the three
phase system.

For the independent concentrations we choose the hydrogen ion $c_{a,H^{+}%
} c,$ concentration the chloride ion $c_{a,Cl^{-}}$ concentration and the sodium ion concentration
$c_{a,Na^{+}}$.   Another assumption consists in taking the chloride and sodium ion concentration as equal. This assumption is to simplify our model.

\section{Mass balance equations for porous medium flow with surface species}

All compounds that are used in the model are built with the atoms $C\left(
4\right)  ,$ $H\left(  1\right)  ,$ $O\left(  -2\right)  ,$ $Ca\left(
2\right)  $, $Na\left(  1\right)  $, $Cl\left(  -1\right)  $ and $C\left(
-4\right)  $. For instance
carbon dioxide consists of one atom of carbon with valence four and two oxygen
atoms with valence minus two. The uncompensated valences result in the charge
of the ion, e.g., $CO_{3}^{-2}$ consists of one atom $C\left(  4\right)  $ and
three oxygen atoms $O\left(  -2\right)  .$ Hence $CO_{3}^{2-}$ has a charge of
minus 2. Organic carbon is denoted by $C\left(  -4\right)  $, leading to for
instance $CH_{4},$ where the four valent carbon is compensated by four
monovalent hydrogens. Two carbons may combine to $C_{2}\left(  -6\right)  $,
and $n$ carbons to $C_{n}\left(  -2n-2\right)  ,$ but we denote all organic
carbon by $C\left(  -4\right)  .$

The concentrations of all species are expressed in term of activities and activity coefficients, where the activity coefficients $\gamma$ depend only on the ionic strength $\mu.$

\subsection{Master species in terms of activities in the aqueous phase}

For the derivation of the mass balance equation we first derive the equations
for the dissolved master species, i.e., $C\left(  4\right)  $, $H\left(
1\right)  $, $O\left(  -2\right)  ,$ $Ca\left(  2\right)  ,$ $Cl\left(
-1\right)  ,$ $Na\left(  1\right)  $ concentrations in the aqueous solution,
which do not include the surface master species.

For convenience we first define the total aqueous inorganic carbon (i.e.,
without hydrocarbon) concentration $C_{a,C\left(  4\right)  }$ by%

\begin{align}
C_{a,C\left(  4\right)  }  &  :=\frac{a_{a,CO_{2}}}{\gamma_{a,CO_{2}}\left(
\mu\right)  }+\frac{a_{a,CO_{3}^{2-}}}{\gamma_{a,CO_{3}^{2-}}\left(
\mu\right)  }+\frac{a_{a,HCO_{3}^{-}}}{\gamma_{a,HCO_{3}^{-}}\left(
\mu\right)  }+\frac{a_{a,CaHCO_{3}^{+}}}{\gamma_{a,CaHCO_{3}^{+}}\left(
\mu\right)  }\nonumber\\
&  +\frac{a_{a,CaCO_{3}}}{\gamma_{a,CaCO_{3}}\left(  \mu\right)  }%
+\frac{a_{a,NaCO_{3}^{-}}}{\gamma_{a,NaCO_{3}^{-}}\left(  \mu\right)  }%
+\frac{a_{a,NaHCO_{3}}}{\gamma_{a,NaHCO_{3}}\left(  \mu\right)  }.
\label{cacarbona2}%
\end{align}

In the same way we define the total aqueous hydrogen concentration by%
\begin{equation}
C_{a,H\left(  1\right)  }:=\left(
\begin{array}
[c]{c}%
2\frac{a_{a,H_{2}O}}{\gamma_{a,H_{2}O}}+\frac{a_{a,H^{+}}}{\gamma_{a,H^{+}%
}\left(  \mu\right)  }+\frac{a_{a,OH^{-}}}{\gamma_{a,OH^{-}}\left(
\mu\right)  }\\
+\frac{a_{a,HCO3^{-}}}{\gamma_{a,OH^{-}}\left(  \mu\right)  }+\frac
{a_{a,CaHCO_{3}^{+}}}{\gamma_{a,CaHCO_{3}^{+}}\left(  \mu\right)  }%
+\frac{a_{a,CaOH^{+}}}{\gamma_{a,CaHCO_{3}^{+}}\left(  \mu\right)  }+\\
\frac{a_{a,NaHCO_{3}}}{\gamma_{a,CaHCO_{3}^{+}}\left(  \mu\right)  }%
\end{array}
\right)  :=2\frac{a_{a,H_{2}O}}{\gamma_{a,H_{2}O}}+\delta C_{a,H\left(
1\right)  }. \label{caahydrogen}%
\end{equation}

\noindent where the concentration of water $c_{a,H_{2}O}=a_{a,H_{2}O}%
/\gamma_{a,H_{2}O}$ is much larger than the other concentrations.

The total aqueous oxygen concentration can be written as%
\begin{equation}
C_{a,O\left(  -2\right)  }:=\left(
\begin{array}
[c]{c}%
\frac{a_{a,H_{2}O}}{\gamma_{a,H_{2}O}}+\frac{3a_{a,CO_{3}^{2-}}}%
{\gamma_{a,CO_{3}^{2-}}\left(  \mu\right)  }+\frac{a_{a,OH^{-}}}%
{\gamma_{a,OH^{-}}\left(  \mu\right)  }+\frac{2a_{a,CO2}}{\gamma
_{a,CO2}\left(  \mu\right)  }\\
+\frac{3a_{a,HCO3^{-}}}{\gamma_{a,HCO3^{-}}\left(  \mu\right)  }
+\frac{3a_{a,CaHCO_{3}^{+}}}{\gamma_{a,CaHCO_{3}^{+}}\left(  \mu\right)
}+\frac{a_{a,CaOH^{+}}}{\gamma_{a,CaOH^{+}}\left(  \mu\right)  }+\\
+\frac{3a_{a,CaCO_{3}}}{\gamma_{a,CaCO_{3}}\left(  \mu\right)  }%
+\frac{3a_{a,NaCO_{3}^{-}}}{\gamma_{a,NaCO_{3}^{-}}\left(  \mu\right)  }
+\frac{3a_{a,NaHCO_{3}}}{\gamma_{a,NaHCO_{3}}\left(  \mu\right)  }%
\end{array}
\right)  :=\frac{a_{a,H_{2}O}}{\gamma_{a,H_{2}O}}+\delta C_{a,O\left(
-2\right)  }. \label{caaoxygen}%
\end{equation}

The total aqueous calcium concentration is
\begin{equation}
C_{a,Ca\left(  2\right)  }:=\frac{a_{a,Ca^{2+}}}{\gamma_{a,Ca^{2+}}\left(
\mu\right)  }+\frac{a_{a,CaHCO_{3}^{+}}}{\gamma_{a,CaHCO_{3}^{+}}\left(
\mu\right)  }+\frac{a_{a,CaOH^{+}}}{\gamma_{a,CaOH^{+}}\left(  \mu\right)
}+\frac{a_{a,CaCO_{3}}}{\gamma_{a,CaCO_{3}}\left(  \mu\right)  }+\beta
_{Ca}/2. \label{caacalcium}%
\end{equation}
where $\beta_{Ca}$ is the equivalent fraction of calcium (see Definition in \cite{appelo2005geochemistry}). 
The total aqueous sodium concentration is
\begin{equation}
C_{a,Na\left(  1\right)  }:=\frac{a_{a,Na^{+}}}{\gamma_{a,Na^{+}}\left(
\mu\right)  }+\frac{a_{a,NaCO_{3}^{-}}}{\gamma_{a,NaCO_{3}^{-}}\left(
\mu \right)}+\frac{a_{a,NaHCO_{3}}}{\gamma_{a,NaHCO_{3}}\left(  \mu\right)
}+\beta_{Na}. \label{caasodium}%
\end{equation}
where $\beta_{Na}$ is the equivalent fraction of sodium.

The total aqueous chloride concentration is
\begin{equation}
C_{a,Cl\left(  -1\right)  }:=\frac{a_{a,Cl^{-}}}{\gamma_{a,Cl^{-}}\left(
\mu\right)  }. \label{caachloride}%
\end{equation}

\subsection{Multiphase mass-balance equations with surface complexes}

\paragraph{Carbon balance}

We can write the mass balance equation for carbon as
\begin{align}
&  \partial_{t}\left(  \varphi C_{a,C\left(  4\right)  }S_{w}+\varphi
c_{o,CO_{2}}S_{o}+\left(  1-\varphi\right)  \left(  c_{r,CaCO_{3}%
}+Cal_{s,CO_{3}^{-}}\right)  \right) \nonumber\\
&  +\partial_{x}u\left(  C_{a,C\left(  4\right)  }f_{w}+c_{o,CO_{2}}%
f_{o}\right) \nonumber\\
&  =\partial_{x}\left(  \varphi\left(  D_{w}S_{w}\partial_{x}C_{a,C\left(
4\right)  }+D_{o}S_{o}\partial_{x}c_{o,CO_{2}}\right)  \right)  +\nonumber\\
&  \partial_{x}\left(  \mathcal{D}C_{a,C\left(  4\right)  }\partial_{x}%
S_{w}\right)  +\partial_{x}\left(  \mathcal{D}c_{o,CO_{2}}\partial_{x}%
S_{o}\right)  , \label{massabalancecarbon}%
\end{align}
where $c_{o,CO_{2}}$ is the concentration of CO$_{2}$ in the oil phase. The
surface master species are arbitrarily chosen to be $oil_{s,N},$
$oil_{w,COO},Cal_{s,O},$ $Cal_{w,CO_{3}}$. This choice uses that the sum of the
species derived from the master species is equal to the number of active
sites, which is considered to be constant for two sets of adsorbed sites on
the oil surface and two sets adsorbed onto the calcite surface. Note that
$S_{o}=1-S_{w}.$ The equilibrium conditions of the sorbed species are taken
from \cite{brady2012surface}.

\paragraph{Hydrogen balance}

In the same way we find for the total hydrogen balance
\begin{align}
&  \partial_{t}\left(  \varphi C_{a,H\left(  1\right)  }S_{w}\right)
+\partial_{t}\varphi\left(  oil_{s,}{}_{^{N-H^{+}}}+oil_{w,COO-H}\right)
S_{o}\nonumber\\
&  +\partial_{t}\left(  1-\varphi\right)  \left(  2Cal_{s,O-2H^{+}%
}+Cal_{s,O-H}+Cal_{w,CO_{3}^{-}-H}\right) \nonumber\\
&  +\partial_{x}\left(  uC_{a,H\left(  1\right)  }f_{w}\right)  +\partial
_{x}u\left(  oil_{s,}{}_{^{N-H^{+}}}+oil_{w,COO-H}\right)  f_{o}=\nonumber\\
&  \partial_{x}\left(  \mathcal{D}C_{a,H\left(  1\right)  }\partial_{x}%
S_{w}\right)  +\partial_{x}\left(  \mathcal{D}\left(  oil_{s,}{}_{^{N-H^{+}}%
}+oil_{w,COO-H}\right)  \partial_{x}S_{o}\right) \nonumber\\
&  +\partial_{x}\left(  \varphi D_{w}S_{w}\partial_{x}C_{a,H\left(  1\right)
}\right)  +\partial_{x}\left(  \varphi D_{o}S_{o}\partial_{x}\left(
oil_{s,}{}_{^{N-H^{+}}}+oil_{w,COO-H}\right)  \right)  ,
\label{massabalancehydrogen}%
\end{align}
where the surface master species are chosen to be $oil_{s,N-},$ $oil_{w,COO-}%
,Cal_{s,O-},$ $Cal_{w,CO_{3}^{-}-}.$

\paragraph{Oxygen balance}

In the same way we find for the total oxygen balance%
\begin{align}
\nonumber
& \partial_{t}\left(  \varphi\left(  C_{a,O\left(  -2\right)  }%
S_{w}+2c_{o,CO_{2}}S_{o}\right)  +\left(  1-\varphi\right)  \left(
3c_{r,CaCO_{3}}+2\left(  Cal_{s,CO_{3}^{-}}\right)  \right)  \right) \\
&+\partial_{x}u\left(  C_{a,O\left(  -2\right)  }f_{w}+2c_{o,CO_{2}}%
f_{o}\right)  
  =\partial_{x}\left(  \mathcal{D}C_{a,O\left(  -2\right)
}\partial_{x}S_{w}\right)  +2\partial_{x}\left(  \mathcal{D}c_{o,CO_{2}}%
S_{o}\partial_{x}S_{o}\right)  \nonumber\\
&+\partial_{x}\varphi\left(  D_{w}S_{w}\partial_{x}C_{a,O\left(  -2\right)
}+2D_{o}S_{o}\partial_{x}c_{o,CO_{2}}\right).  \label{massabalanceoxygen}%
\end{align}

\paragraph{Calcium balance}

For the total calcium we find

\begingroup\everymath{\scriptstyle}
\begin{align}
&  \mathtt{\partial_{t}\left(  \varphi S_{w}C_{a,Ca\left(  2\right)  }+\varphi
S_{o}\left(  oil_{w,COO-Ca^{+}}\right)  +\left(  1-\varphi\right)  \left(
c_{r,CaCO_{3}}+Cal_{w,CO_{3}-Ca^{+}}\right)  \right)  +\partial_{t}\beta}%
_{Ca}\nonumber\\
&  \partial_{x}u\left(  C_{a,Ca\left(  2\right)  }f_{w}\right)  +\partial
_{x}u\left(  oil_{w,COO-Ca^{+}}\right)  f_{o}=\partial_{x}\left(
\mathcal{D}C_{a,Ca\left(  2\right)  }\partial_{x}S_{w}\right) \nonumber\\
&  +\partial
_{x}\left(  \mathcal{D}oil_{w,COO-Ca^{+}}\partial_{x}S_{o}\right)  +\partial_{x}\left(  \varphi D_{w}S_{w}\partial_{x}C_{a,Ca\left(  2\right)
}\right)  +\partial_{x}\left(  \varphi D_{o}S_{o}\partial_{x}oil_{w,COO-Ca^{+}%
}\right).  \label{massabalancecalcium}%
\end{align}
\endgroup

\paragraph{Sodium, Chlorine and total oil equation}

For the sodium equation we obtain
\begin{equation}
\partial_{t}\left(  \varphi S_{w}C_{a,Na\left(  1\right)  }\right)
+\partial_{t}\beta_{Na}+\partial_{x}\left(  uC_{a,Na\left(  1\right)  }%
f_{w}\right)  =\partial_{x}\left(  \mathcal{D}C_{a,Na\left(  1\right)
}\partial_{x}S_{w}\right)  +\partial_{x}\left(  \varphi D_{w}S_{w}\partial
_{x}C_{a,Na\left(  1\right)  }\right).  \label{massabalancesodium}%
\end{equation}
In the same way we obtain for the first order terms of the chlorine equation
\begin{equation}
\partial_{t}\left(\varphi S_{w}C_{a,Cl\left(  -1\right)  }\right)+\partial_{x}%
\left( u C_{a,Cl\left(  -1\right)  }f_{w}\right)  =\partial_{x}\left( 
\mathcal{D}C_{a,Cl\left(  -1\right)  }\partial_{x}S_{w}\right)  +\partial
_{x}\left(  \varphi D_{w}S_{w}\partial_{x}C_{a,Cl\left(  -1\right)
}\right).  \label{massabalancechloride}%
\end{equation}
For the total oil, e.g., heptane we retain the first order terms%
\begin{equation}
\partial_{t}\left(  \varphi S_{o}c_{o,C\left(  -4\right)  }\right)
+\partial_{x}\left(  u c_{o,C\left(  -4\right)  }f_{o}\right)
=\partial_{x}\left(  \mathcal{D}c_{o,C\left(  -4\right)  }\partial_{x}%
S_{o}\right)  +\partial_{x}\left(  \varphi D_{o}S_{o}\partial
_{x}c_{o,C\left(  -4\right)  }\right). \label{massabalanceoil}%
\end{equation}
The oil concentration can be obtained from the dissolved carbon dioxide
concentration $c_{o,CO_{2}}$ (proportional to $c_{a,CO_{2}}$) with the EOS
\begin{equation}
\frac{c_{o,C\left(  -4\right)  }}{c_{O,C\left(  -4\right)  }}+\frac
{c_{o,CO_{2}}}{c_{O,CO2}}=1. \label{oilEOS}%
\end{equation}
\subsection{System of conservation laws}

Based on Gibbs rule and assuming that sodium and chloride concentration are similar
we consider four conservation laws, namely for total carbon, hydrogen$-$oxygen, chloride and decane. Each one consists of four terms, i.e. accumulation, convection, molecular diffusion and capillary diffusion. We neglect diffusion and capillarity effects.
	
We can write the mass balance equation for carbon as
\begin{align}
	&  \partial_{t}\left(  \varphi C_{a,C\left(  4\right)  }S_{w}+\varphi
	c_{o,CO_{2}}S_{o}+\left(  1-\varphi\right)  \left( Cal_{s,CO_{3}^{-}}\right)  \right) \nonumber\\
	&  +\partial_{x}\left (u\left(  C_{a,C\left(  4\right)  }f_{w}+c_{o,CO_{2}}%
	f_{o}\right)\right )=0.
   \label{massabalancecarbon1}%
\end{align}
For the total oil, e.g., heptane we obtain
\begin{equation}
	\partial_{t}\left(  \varphi S_{o}c_{o,C\left(  -4\right)  }\right)
	+\partial_{x}\left(  u c_{o,C\left(  -4\right)  }f_{o}\right)
	=0  . \label{massabalanceoil1}%
\end{equation}
In the same way we obtain the chlorine equation
\begin{equation}
	\partial_{t}\varphi S_{w}C_{a,Cl\left(  -1\right)  }+\partial_{x}%
	 \left( u  C_{a,Cl\left(  -1\right)  }f_{w}\right)  =0.  \label{massabalancechloride1}%
\end{equation}

To remove one balance equation, we combine the hydrogen and oxygen balance equations to a single equation, in such a way that the water concentration is eliminated. We do so because the water concentration is much higher than the other concentrations.

We subtract twice the oxygen equation from the hydrogen equation and obtain
after substitution of Eq. (\ref{massabalancehydrogen}) and Eq.
(\ref{massabalanceoxygen})%
\begin{align}
	&  \partial_{t}\left(  2\delta C_{a,O\left(
		-2\right)  }-\delta C_{a,H\left(  1\right)  }\right)  S_{w}-\partial_{t}\varphi\left(  oil_{s,}{}_{^{N-H^{+}}%
	}+oil_{w,COO-H}\right)  S_{o}\nonumber\\
	&  -\partial_{t}\left(  1-\varphi\right)  \left(  2Cal_{s,O-2H^{+}%
	}+Cal_{s,O-H}+Cal_{w,CO_{3}^{-}-H}\right) \nonumber\\
	&  2\partial_{t}\left(  \varphi\left(  2c_{o,CO_{2}}S_{o}\right)  +\left(
	1-\varphi\right)  \left(  3c_{r,CaCO_{3}}+2\left(  Cal_{s,CO_{3}^{-}}\right)
	\right)  \right) \nonumber\\
	&  +\partial_{x}u\left( 4c_{o,CO_{2}}- oil_{s,}{}_{^{N-H^{+}}}-oil_{w,COO-H}\right)
	f_{o}+\partial_{x}u\left(  \left( 2 \delta
	C_{a,O\left( -2\right) }- \delta C_{a,H\left(  1\right)  }\right)  f_{w} \right )
	=0. \label{masbaloxminhyd1}%
\end{align}
In more compact way the system of conservation laws \eqref{massabalancecarbon1} and \eqref{masbaloxminhyd1} can be written
\begin{align}
 \partial_{t}\left(\varphi\rho_{w1} S_{w}+\varphi\rho_{o1} S_{o}+\rho_{r1}  \right) 
 +\partial_{x}\left(u\left(\rho_{w1} f_{w}+\rho_{o1}
f_{o}\right)\right)=0 
, \label{massabalancecarbon1a1}%
\end{align}
\begin{align}
 \partial_{t}\left(\varphi\rho_{o2} S_{o} \right) 
+\partial_{x}\left( u\left( \rho_{o2}
f_{o}\right)\right)=0 
, \label{massabalancecarbon1a123}%
\end{align}
\begin{align}
\partial_{t}\left(\varphi\rho_{w3} S_{w} \right) 
+\partial_{x}\left(u\left( \rho_{w3}
f_{w}\right)\right)=0 
, \label{massabalancecarbon1a12}%
\end{align}
\begin{align}
\partial_{t}\left(\varphi\rho_{w4} S_{w}+\varphi\rho_{o4} S_{o}+\rho_{r4}  \right) 
+\partial_{x}\left( u\left( \rho_{w4} f_{w}+\rho_{o4}
f_{o}\right)\right)=0 
, \label{massabalancecarbon1a14}%
\end{align}
where the coefficient functions are defined by
\begin{eqnarray}
\label{ereq1}
&\rho_{w1}= C_{a,C\left(  4\right)  },~\rho_{o1}=c_{o,CO_{2}},~\rho_{r1}= (1-\varphi)\left( Cal_{s,CO_{3}^{-}}\right),\\
&\rho_{o2}= C_{a,C\left(  -4\right)  }, ~~\rho_{w3}= C_{a,Cl},\\
&\rho_{w4}=\left( 2\delta C_{a,O\left(
	-2\right)  }- \delta C_{a,H\left(  1\right)  }\right),~\rho_{o4}= \left ( 4c_{o,CO_{2}} -\left(  oil_{s,}{}_{^{N-H^{+}}%
}+oil_{w,COO-H}\right)\right ),\\
\label{ereq2}
&\rho_{r4}= (1-\varphi)\left( 4 Cal_{s,CO_{3}^{-}}-\left(  2Cal_{s,O-2H^{+}%
}+Cal_{s,O-H}+Cal_{w,CO_{3}^{-}-H}\right)\right).
\end{eqnarray}
where $\rho_{wi}$, $\rho_{oi}$ and  $\rho_{ri}$ depend on $pH$ and $[Cl]$. 
It is possible to verify that when the chemical surface complexes are not present then system \eqref{massabalancecarbon1a1}-\eqref{massabalancecarbon1a14} is reduced to the system studied in \cite{alvarez2018analytical}.

\subsection{Fractional flow}

The fractional flows for water and oil are saturation-dependent functions defined as follows. We denote 
$S_e(S_w)=(S_w-S_{wr})/(1-S_{wr}-S_{or}),~\text{for}~
S_w \geq S_{wr}~\text{and}~S_e=0~ \text{for}$ $S_w < S_{wr}; k_{rw}(S_w)=k_w(S_e(S_w))^{n_w}~~ \text{ and }~~ k_{ro}(S_w)=k_o(1-S_e(S_w))^{n_o}$ ($S_o=1-S_w$) (\cite{etemadi2017modelling}). Here $n_w$ and $n_o$ are the Corey exponents. The parameters $k_w$ and $k_o$ are end point relative permeabilities of water and oil
phases, respectively. 

The water viscosity is taken as $\mu_w=0.001$ and the oil viscosity
as $\mu_o=0.002$ when they are constant; then we have the fractional flow functions for the aqueous and oleic phases
\begin{equation}
f_w(S_w)=\frac{k_{rw}(S_w)/\mu_w}{(k_{rw}(S_w)/\mu_w+k_{ro}(1-S_w)/\mu_o)},\quad \text{and} \quad f_o(S_w)=1-f_w(S_w), 
\label{eq1a}
\end{equation}
where the water and oil permeabilities $k_{rw}(S_w)$ and $k_{ro}(S_o)$ 
are expressed in terms of their saturations; $\mu_w$ and $\mu_o$ are 
the viscosities of the aqueous and oleic phases. We disregard capillarity and diffusive 
effects.

In this work we consider the relation between relative permeability and low salinity from the ideas
described in \cite{jerauld2008modeling}. Such a relation is based on the
definition of the weight parameter $\theta$ to modify rock and fluid properties
regarding salinity level.  This model
considers salt as an aqueous component that can be transported and
traced. In this relation the relative permeability
functions depend on water saturation $S_w$ and salinity. This relationship is defined by the upper ($HL$)
and lower ($LS$) limits for salinity called high and low salinity limits,
respectively. To do that Corey's coefficients are adjusted for low and high salt concentration.

The weighting function $\theta$ is introduced by a linear
relationship among two values for residual oil saturation corresponding to high and low salinity limits
\begin{equation}
    \theta=\frac{S_{or}-S_{orw}^{LS}}{S_{orw}^{HL}-S_{orw}^{LS}},
    \label{thetae}
\end{equation}
where the parameter $\theta$ is used for interpolating between preset high and low salinity
curves for relative permeability and capillary pressure
\begin{equation}
    K_{wi}(S_w)=\theta K_{ri}^{HL}(S_w)+(1-\theta)K_{ri}^{LS}(S_w),
    \label{thetae1}
\end{equation}
where $i=w,o$ for water and oil permeability, respectively. For high salinity regime we take
$k_w=0.25$, $k_o=0.5$, $n_w=3$ and $n_o=2$, while for low salinity the values $k_w=0.41$, $k_o=0.5$, $n_w=3$ and $n_o=2$ are used. Furthermore, in this study, the upper limit used for salinity corresponds to maximum residual oil saturation ($S_{or}^{H}=0.35$). The lower limit corresponds to minimum residual oil saturation
($S_{or}^{L}=0.15$).

\section{Surface complexation modeling}

In this study,  we use SCM to predict the wettability alteration of minerals through TBP. Details of the  procedure employed here can be found in \cite{ginn2020effects,mehdiyev2022surface}.
The geochemistry solver PHREEQC was utilized to calculate the concentrations of the complexes. This program uses as input the properties of formation water and oil components of the crude oil. The properties of the minerals for several formation waters and for two polar oil components in the crude oil acid (e.g., COOH and NH$^+$) are used here. We take the input data described in \cite{erzuah2019wettability}. Input properties of Oil are depicted in Table \ref{tab1} and \ref{tab2}.
\begin{table}[h]
\begin{tabular}{|l|l|l|l|}
\hline
Oil & Density (g/cm³) at 20$^{o}$C & TAN (mg KOH/g oil) & TBN (mg KOH/g oil) \\ \hline
1     & 0.86                                       & 0.1                & 1.9                \\ \hline
2     & 0.9                                        & 0.38               & 2.3                \\ \hline
\end{tabular}
\caption{\label{tab1}Properties of the minerals and the polar oil components.}
\end{table}

The total acidic number (TAN) and the total base number (TBN) are used to calculate the oil site density using the formulas (see \cite{bonto2019overview}) 
\begin{equation}
    N_{S,COOH}=0.602 \times 10^6 \frac{TAN}{1000~ a_{oil}~ MW_{KOH}},
\end{equation}

\begin{equation}
    N_{S,NH^+}=0.602 \times 10^6 \frac{TBN}{1000~ a_{oil}~ MW_{KOH}},
\end{equation}
where $MW_{KOH}=56.1 g/mol$ molecular weight of potassium hydroxide.  Here $a_{oil}$ [$m^2/g$] denotes the specific area of oil, which we assume to be the same as its respective carbonate minerals in aqueous solutions taken from \cite{wolthers2008surface}.
\begin{table}[h]
\begin{tabular}{|l|l|l|l|}
\hline
Surface & Site Density (site/$nm^2$) & Surface Area ($m^2$/g) & Mass (g) \\ \hline
Calcite & 4.9                     & 2.0                 & 0.2      \\ \hline
Oil 1=COOH    & 0.54                    & 2.0                 & 2.59     \\ \hline
Oil 1=NH+     & 10.2                    & 2.0                 & 2.59     \\ \hline
Oil 2=COOH    & 2.04                    & 2.0                 & 2.71     \\ \hline
Oil 2=NH+     & 12.34                   & 2.0                 & 2.71     \\ \hline
\end{tabular}
\caption{\label{tab2}Surface density and area of the two types of oil.}
\end{table}

Input data sets corresponding to salt are chosen for several formation waters where the ion concentrations of $Na^{2+}$ and $Cl^{-}$ varies. The data are taken to yield the coefficients of system \eqref{massabalancecarbon1a1}-\eqref{massabalancecarbon1a14}, which depend on the ion concentration chosen in this work to represent the variability of state space, i.e., $pH$ and $Cl^{-}$. Since we assume that the ion concentrations of $Na^{2+}$ and $Cl^{-}$ are similar in the simulations, these ion concentrations vary from 30 to 1290 mmol/kgw. In turn, the ion concentrations of $Ca^{2+}$, $Mg^{2+}$ and $SO_4^{2-}$ are taken as constant. The $pH$ is taken to vary from 2.7 to 8. Carbon is also taken as a constant in this study. The data are summarized in Table \ref{tab3}.
\begin{table}[h]
\begin{tabular}{|l|l|}
\hline
Ion   & Synthetic injected water( mmol/kgw) \\ \hline
$Na^{+}$    & 30-1290                              \\ \hline
$Mg^{2+}$    & 40                                   \\ \hline
$Ca^{2+}$     & 20                                   \\ \hline
$Cl^{-}$     & 30-1290                              \\ \hline
$SO_{4}^{2-}$  & 20                                   \\ \hline
\end{tabular}
\caption{\label{tab3}Formation waters where chloride and sodium ion concentrations vary. Other minerals are fixed.}
\end{table}
\subsection{Regression formulas Matlab, Eureqa and PHREEQC program}
The expressions for the coefficients $\rho_{wi},~\rho_{oi}$ and
$\rho_{ri}$ ($i=1,\ldots,4$) in system \eqref{massabalancecarbon1a1}-\eqref{massabalancecarbon1a14} are obtained  using the ion concentrations of the complexes with formulas given in \ref{ereq1}-\ref{ereq2}. These coefficients can also be determined by allowing the other minerals as $Ca^{2+}$ and $Mg^{2+}$ to vary as well. In such a case, and in accordance with the generalized Gibbs rule, it is necessary to add other equations to the model and to increase the number of degrees of freedom. This useful procedure can be used to study enhanced oil recovery by Smart Water injection, which we will study in future work with the methodology explained here.

Processing of the PHREEQC output data is done with MATLAB program, by using the curve fitting tool.  
\begin{figure}[ht]
	\centering
	\includegraphics[width=6cm]{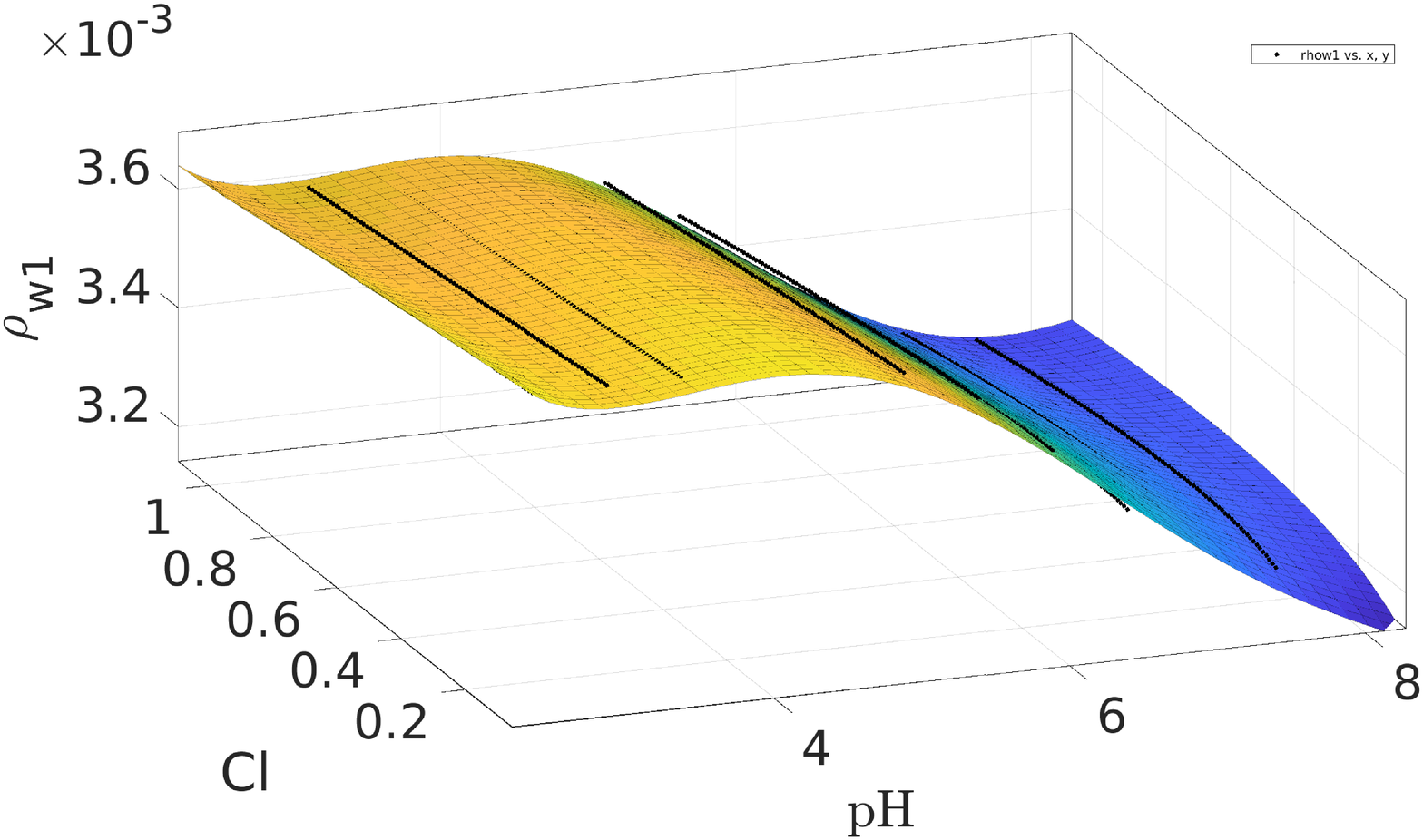}
   \includegraphics[width=6cm]{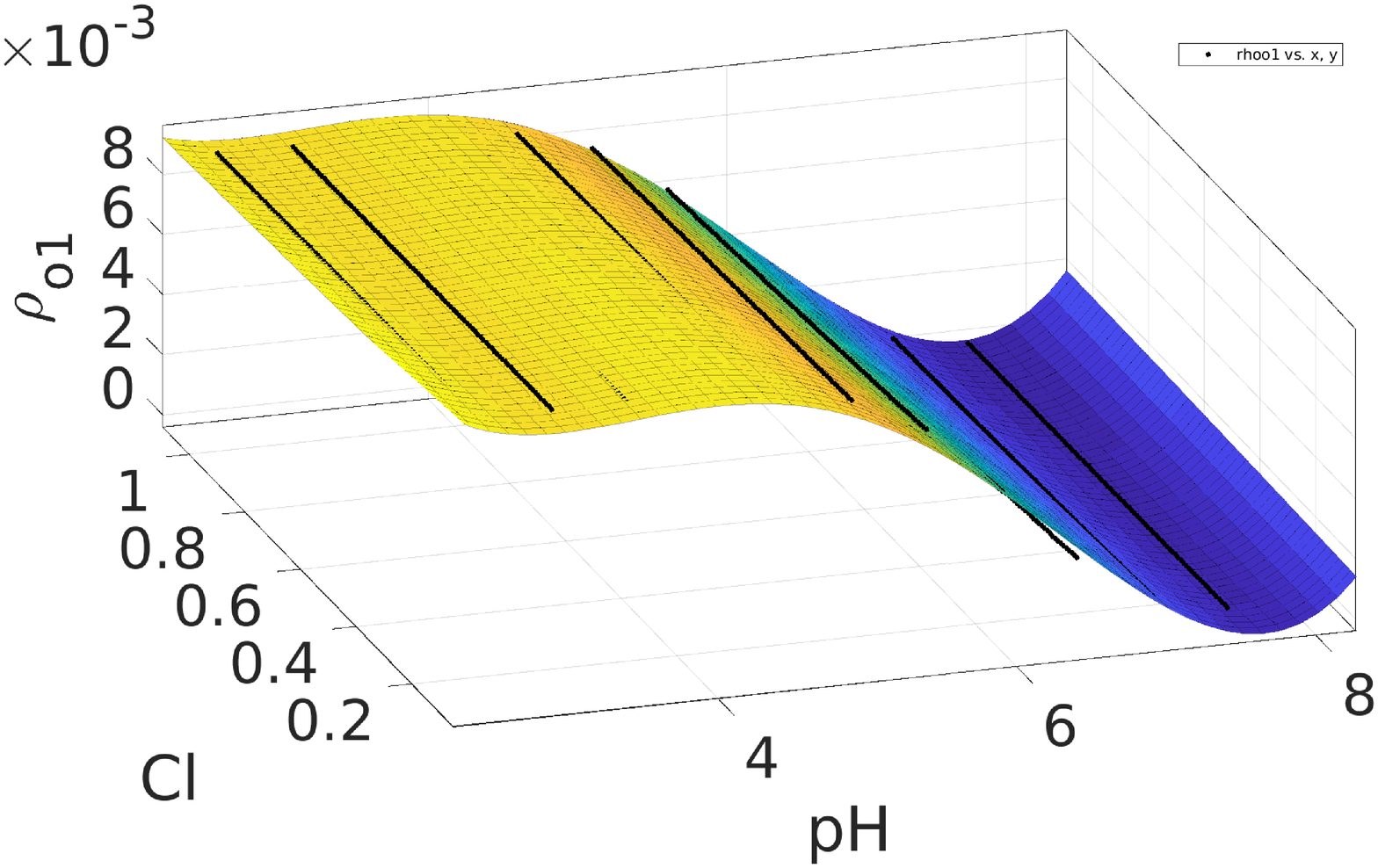}
	\caption{Coefficients $\rho_{w1}$(left) and $\rho_{o1}$(right).}  \label{fig:xray1}
\end{figure}
\begin{figure}[ht]
	\centering
	\includegraphics[width=6cm]{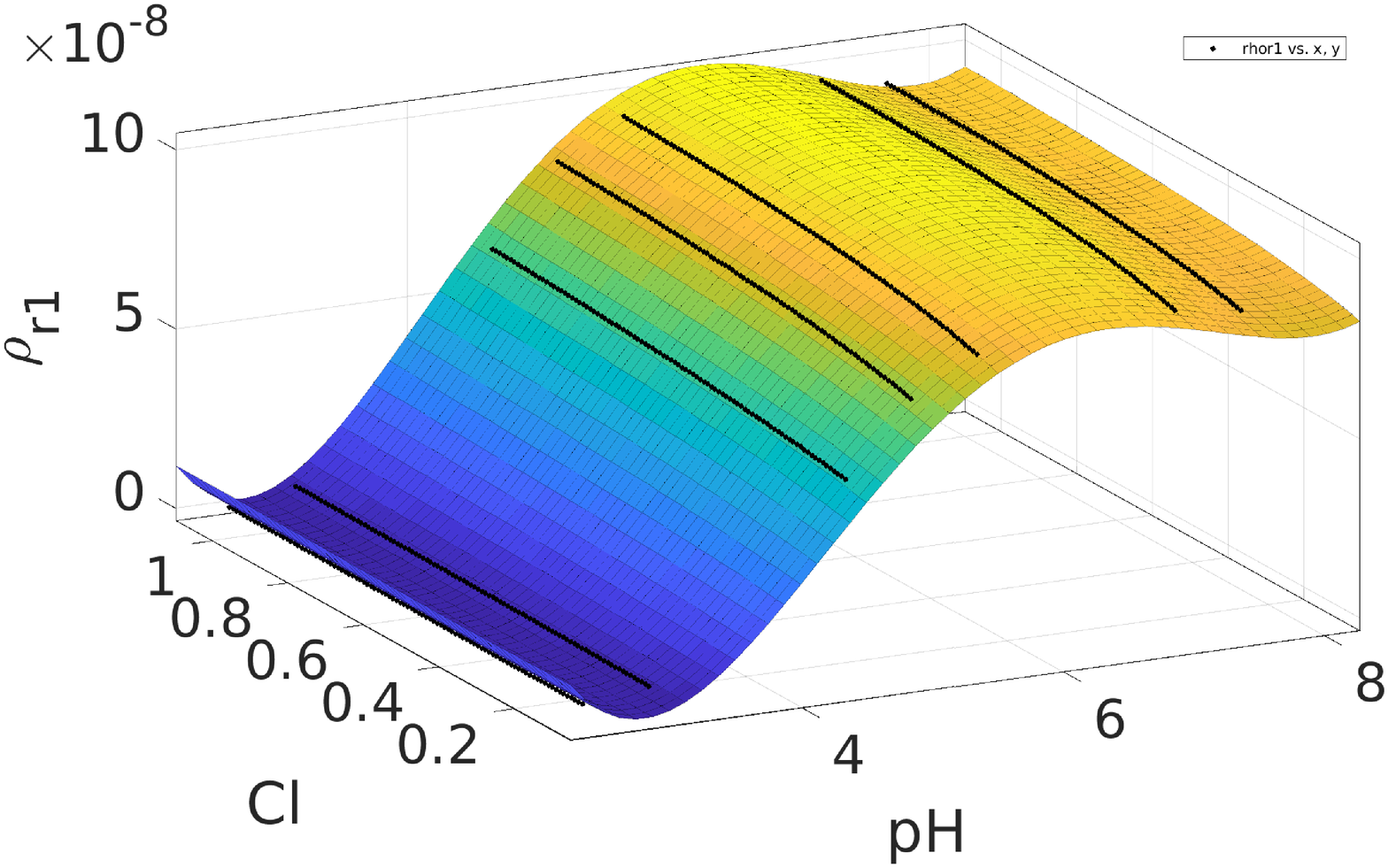}
	\includegraphics[width=6cm]{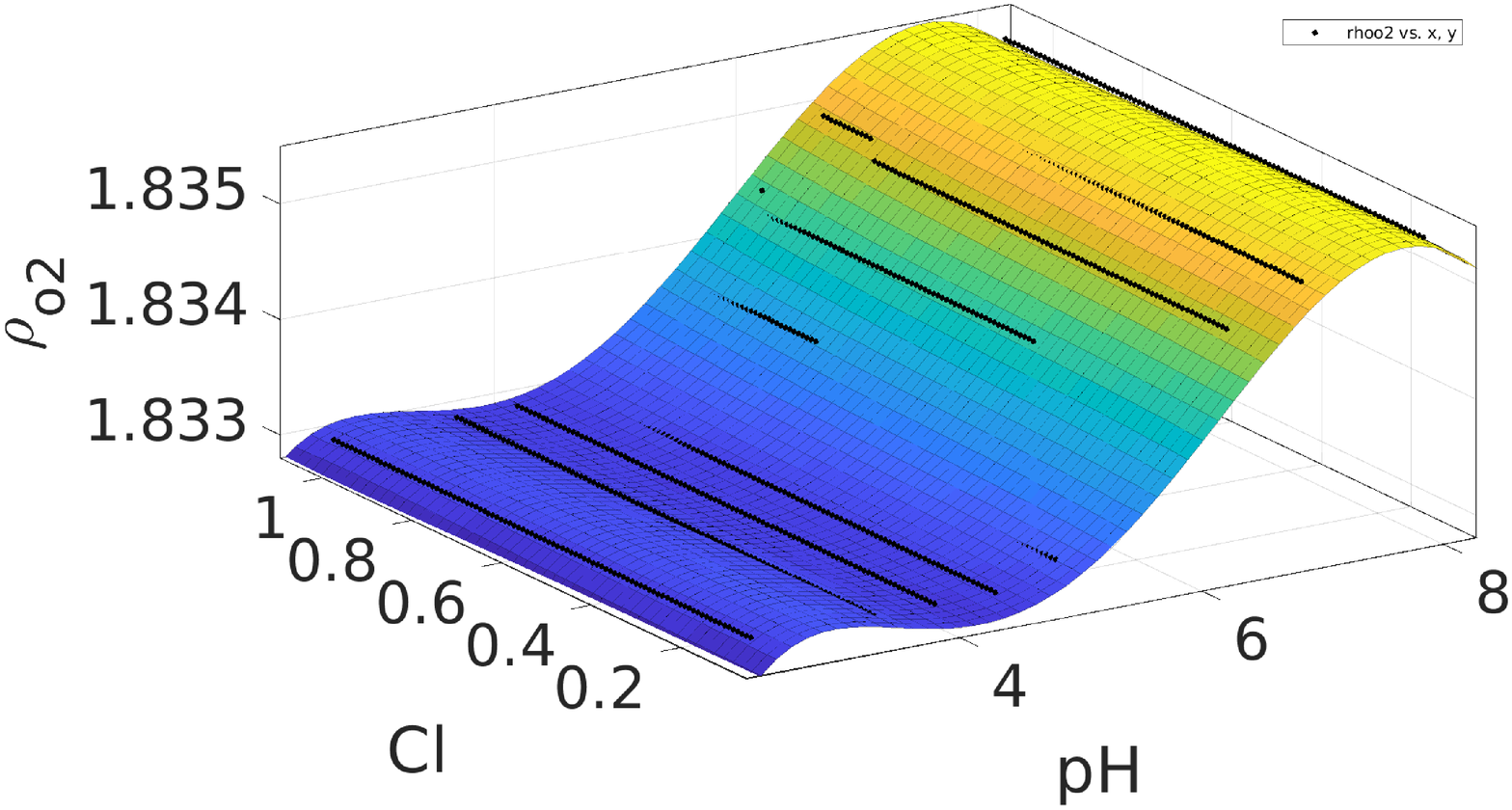}
	\caption{Coefficients $\rho_{r1}$(left) and $\rho_{o2}$(right).}  \label{fig:xray2}
\end{figure}
\begin{figure}[ht]
	\centering
	\includegraphics[width=6cm]{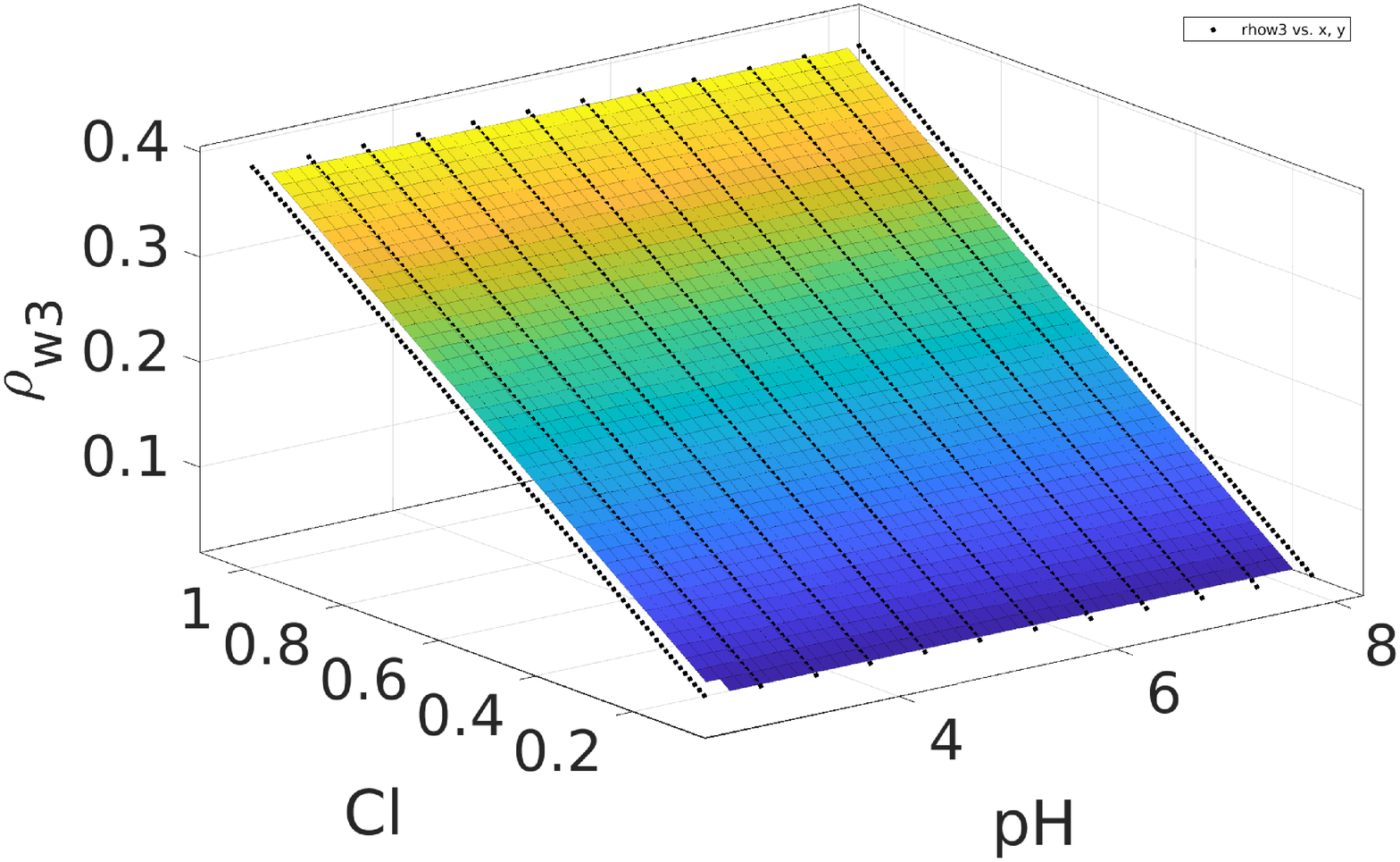}
   \includegraphics[width=6cm]{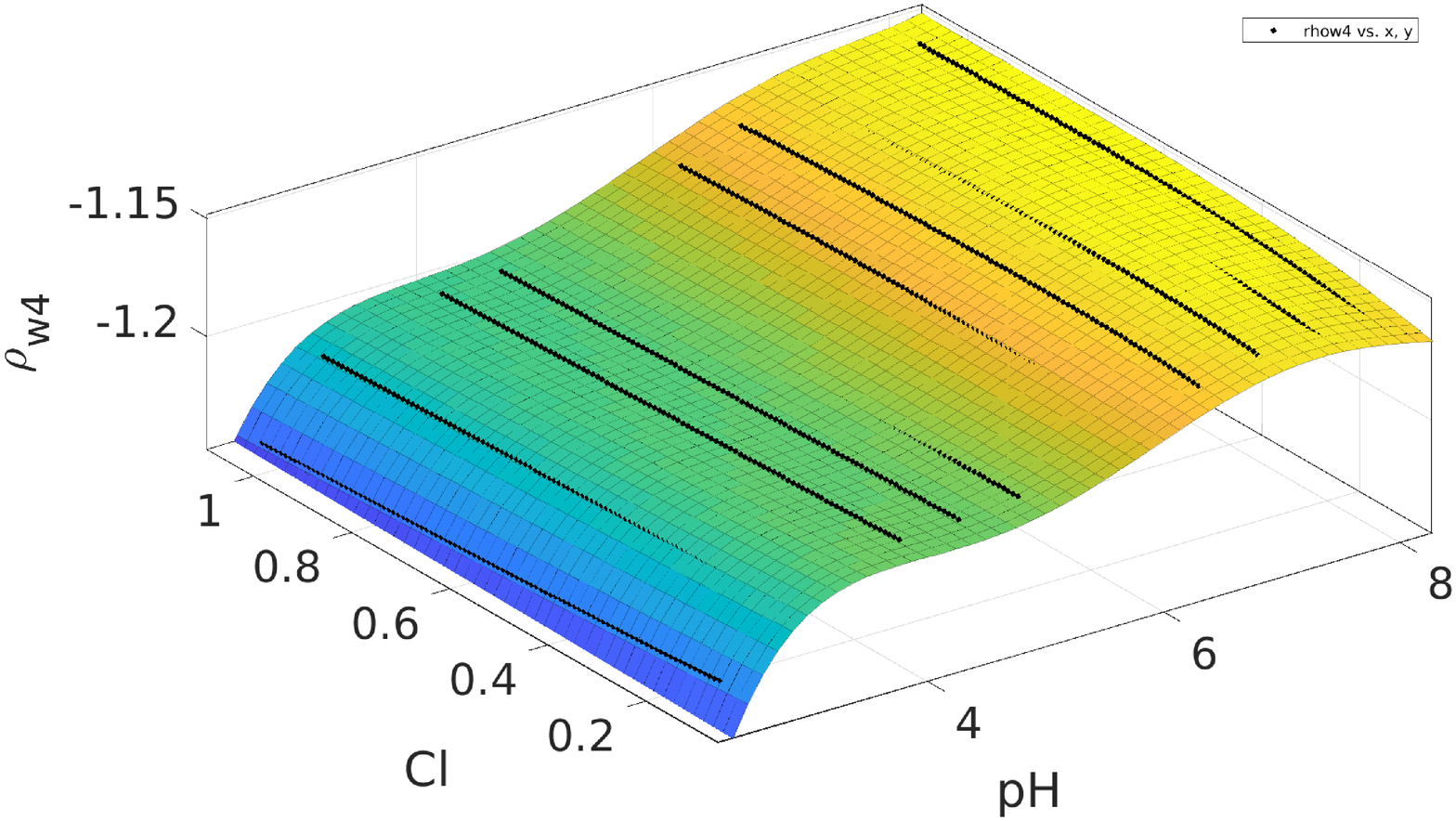}
	\caption{Coefficients $\rho_{w3}$(left) and $\rho_{w4}$(right).}  \label{fig:xray3}
\end{figure}
\begin{figure}[ht]
	\centering
	\includegraphics[width=6cm]{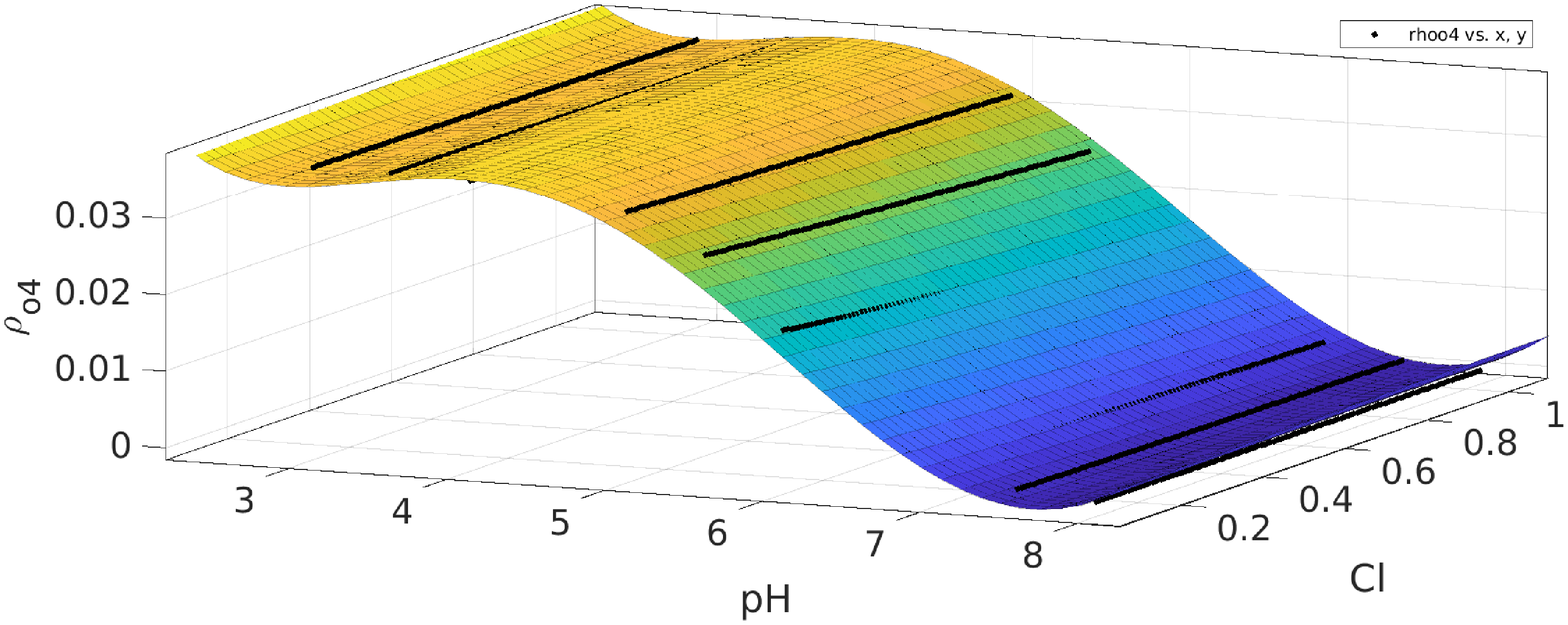}
	\includegraphics[width=6cm]{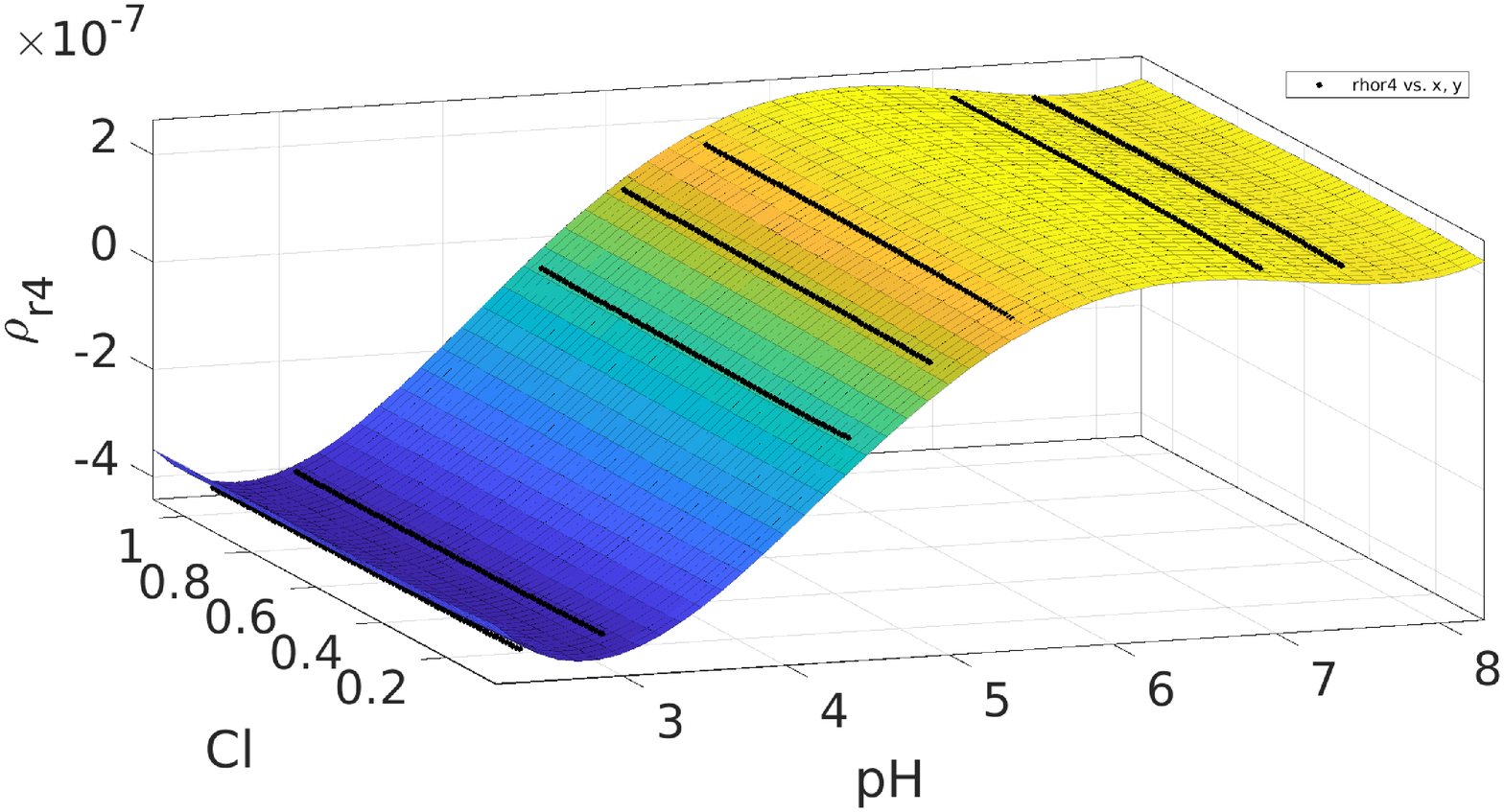}
	\caption{Coefficients $\rho_{o4}$(left) and $\rho_{r4}$(right).}  \label{fig:xray4}
\end{figure}

Denoting the ion concentration of hydrogen and chloride by $x=pH$ and $y=[Cl]$ the formula of the coefficients for the case of Oil $1$ are
\begin{align}
\label{coefeq1}
\nonumber
&\rho_{w1}= -8.247\cdot10^{-7}\,x^5+8.841\cdot10^{-9}\,x^4\,y+2.472\cdot10^{-5}\,x^4+9.621\cdot10^{-9}\,x^3\,y^2\\
\nonumber
&+3.387\cdot10^{-9}\,x^3\,y-2.803\cdot10^{-4}\,x^3
+9.781\cdot10^{-6}\,x^2\,y^3-2.523\cdot10^{-5}\,x^2\,y^2-\\
\nonumber
&2.843\cdot10^{-5}\,x^2\,y+0.0015\,x^2-5.102\cdot10^{-5}\,x\,y^4+
4.901\cdot10^{-5}\,x\,y^3\\
\nonumber
&+8.39\cdot10^{-5}\,x\,y^2+9.35\cdot10^{-5}\,x\,y-0.0036\,x
+1.49\cdot10^{-4}\,y^5-2.383\cdot10^{-4}\,y^4\\
&+1.341\cdot10^{-4}\,y^3-1.758\cdot10^{-4}\,y^2-1.601\cdot10^{-4}\,y+0.007,
\end{align}
\begin{align}
\label{coefeq21}
\nonumber
&\rho_{o1}=- 1.575\cdot10^{-5}x^4y + 1.01\cdot10^{-4}x^4 - 1.83\cdot10^{-5}x^3*y^2 + 3.95\cdot10^{-4}x^3y - \\
\nonumber
&0.002x^3 + 3.9010\cdot10^{-6}x^2y^3 + 2.644\cdot10^{-4}x^2y^2 - 0.0034x^2y + 0.0132x^2, \\
\nonumber
&+ 1.747\cdot10^{-5}xy^4 - 1.0440\cdot10^{-4}xy^3 - 0.0011xy^2 + 0.0115xy - 0.0369x + \\
&4.917\cdot10^{-4}y^5 - 0.0014y^4 + 0.0015y^3 + 7.16\cdot10^{-4}y^2 - 0.0134y + 0.046,
\end{align}
\begin{align}
\label{coefeq21a}
\nonumber
&\rho_{r1}=- 1.056\cdot10^{-10}x^5 + 1.526\cdot10^{-10}x^4y + 3.667\cdot10^{-9}x^4 + 3.284
\cdot10^{-10}x^3y^2,\\
\nonumber
& - 3.606\cdot10^{-9}x^3y - 4.844
\cdot10^{-8}x^3 - 6.403\cdot10^{-10}x^2y^3 - 3.359
\cdot10^{-9}x^2y^2, \\
\nonumber
&+ 2.797\cdot10^{-8}x^2y + 2.980
\cdot10^{-7}x^2 - 3.184\cdot10^{-9}xy^4 + 1.668
\cdot10^{-8}xy^3, \\
\nonumber
&- 6.75\cdot10^{-9}xy^2 - 7.32
\cdot10^{-8}xy - 8.22\cdot10^{-7}x + 1.03
\cdot10^{-8}y^4 - 4.66\cdot10^{-8}y^3 \\
&+ 4.384\cdot10^{-8}y^2 + 5.35\cdot10^{-8}y + 8.19\cdot10^{-7},
\end{align}
\begin{align}
\label{coefeq21b}
\nonumber
&\rho_{o2}=3.742\cdot10^{-6}x^5 + 4.911\cdot10^{-6}x^4*y - 1.278\cdot10^{-4}x^4 - 4.653\cdot10^{-6}x^3y^2,\\
\nonumber
& - 1.09\cdot10^{-4}x^3y + 0.0016x^3 - 3.25\cdot10^{-5}x^2y^3 + 1.2\cdot10^{-4}x^2y^2 + 8.35\cdot10^{-4}x^2y, \\
\nonumber
&- 0.0087*x^2 + 9.718\cdot10^{-5}xy^4 + 1.862\cdot10^{-4}xy^3 - 7.897\cdot10^{-4}xy^2, \\
\nonumber
&- 0.0025xy + 0.022x - 6.073
\cdot10^{-4}y^4 + 4.18\cdot10^{-4}y^3 + 9.108\cdot10^{-4}y^2 \\
&+ 0.0027*y + 1.812,
 \end{align}
 \begin{equation}
 \rho_{w3}=0.37y,
 \end{equation}
 \begin{align}
\label{coefeq21c}
\nonumber
&\rho_{w4}=2.61\cdot10^{-4}x^5 + 1.67\cdot10^{-4}x^4y - 0.0079x^4 + 2.78\cdot10^{-5}x^3y^2, \\
\nonumber
&- 0.003x^3y + 0.0914x^3 + 8.04\cdot10^{-4}x^2y^3 - 0.0025x^2y^2 + 0.025x^2y - 0.51x^2,\\
\nonumber
& - 0.0038xy^3 + 0.0113xy^2 - 0.0744xy + 1.3520x + 0.0021y^3 - 0.0072y^2 \\
&+ 0.0616y - 2.5890,
\end{align}
\begin{align}
\label{coefeq21d}
\nonumber
&\rho_{o4}=- 5.701\cdot10^{-5}x^5 - 6.681\cdot10^{-5}x^4*y + 0.0019*x^4 - 7.327\cdot10^{-5}x^3y^2\\
\nonumber
&  + 0.0017x^3y - 0.0231x^3 + 9.061\cdot10^{-6}x^2y^3 + 0.0011x^2y^2 - 0.0142x^2y \\
\nonumber
& + 0.1273x^2 + 2.365\cdot10^{-5}xy^4 - 1.965
\cdot10^{-4}xy^3 - 0.0046xy^2 + 0.0483xy\\
& - 0.322x + 9.37\cdot10^{-4}y^5 - 0.0023y^4 + 0.002y^3 + 0.006y^2 - 0.057y + 0.34,
\end{align}
\begin{align}
\label{coefeq21e}
\nonumber
&\rho_{r4}=5.86\cdot10^{-9}x^4 - 3.762\cdot10^{-10}x^3y - 1.33\cdot10^{-7}x^3 + 1.228\cdot10^{-10}x^2y^2 \\
\nonumber
&+ 3.162\cdot10^{-9}x^2y + 1.055\cdot10^{-6}x^2 + 8.124\cdot10^{-9}xy^3 - 2.369\cdot10^{-8}xy^2 \\
\nonumber
&+ 1.87
\cdot10^{-8}xy - 3.282\cdot10^{-6}x - 
4.032\cdot10^{-8}y^3 + 1.093\cdot10^{-7}y^2 \\
&- 1.056\cdot10^{-7}y + 3.072\cdot10^{-6}.
 \end{align}

The form of the function described for the above coefficients is the same for the two types of oil used in this study. These formulas show how much the pH affects the coefficients (see Figures \ref{fig:xray1}-\ref{fig:xray4}). In all coefficients the significant changes happen around $pH=5$.

The coefficients in \eqref{coefeq1}-\eqref{coefeq21e} are used as input to solve the system of conservation laws  \eqref{massabalancecarbon1a1}-\eqref{massabalancecarbon1a14}. First we use the numerical solution from the solver COMSOL, which is based on the finite element method. Second, we obtain the solution by the in house Riemann solver RPNfilho. To obtain a numerical solution we provide the coefficients and explore a set of parameters which permit to obtain stable solutions.

\subsection{Wettability estimation}

As suggested in \cite{erzuah2019wettability}, we assume that wettability is proportional to the Total Bound product (TBP). TBP estimation is based on the quantification with SCM of the attractive electrostatic forces existing between the rock-brine and the oil-brine interfaces. 
These forces are represented by Bound Product (BP), which is the
product of the mole fractions of the oil ($O_i$) and mineral ($m_i$) sites with unlike charges \cite{erzuah2019wettability}
and it is given by the Equation:
\begin{equation}
    BP=O_i*m_i
\end{equation}
For a given mineral or rock, the sum of all the BP is the Total Bond Product (TBP) \cite{erzuah2019wettability}.
The SCM using PHREEQC solver predicts the oil adhesion onto a mineral or rock surface using the electrostatic
pair linkages existing between the mineral–brine and oil–brine interfaces. It has been reported in the literature that the higher TBP corresponds to the
higher tendency for oil to be adsorbed onto the surface and vice-versa. 
The TBP can be written as:
\begin{align}
\nonumber
 &TBP_1=(oil_{w,COO-})*(Cal_{s,CaOH2+}+Cal_{w,CO3Ca+}+
Ca_{w,CO3Mg+}),  \\ 
\nonumber
& TBP_2=(oil_{s,NH+})*(Cal_{w,CO3-}+Ca,_{s,CaO-}+Cal_{s,CaCO3-}+
\nonumber
Cal_{s,CaSO4-}), \\  
\nonumber
& TBP_3=oil_{w,COOCa+}*(Cal_{w,CO3-}+Cal_{s,CaO-}+
Cal_{s,CaCO3-}+Cal_{s,CaSO4-}),  \\
\nonumber
&  TBP_4=oil_{w,COOMg+}*(Cal_{w,CO3-}+Cal_{s,CaO-}+Cal_{s,CaCO3-}
+Cal_{s,CaSO4-}), \\ 
\nonumber
& TBP=TBP_1+TBP_2+TBP_3+TBP_4.
\end{align}
The complexes of oil and rock are calculated with the PHREEQC solver with input values given in Tables \ref{tab1}-\ref{tab3}.

\begin{figure}[h]
	\centering
	\includegraphics[width=10cm]{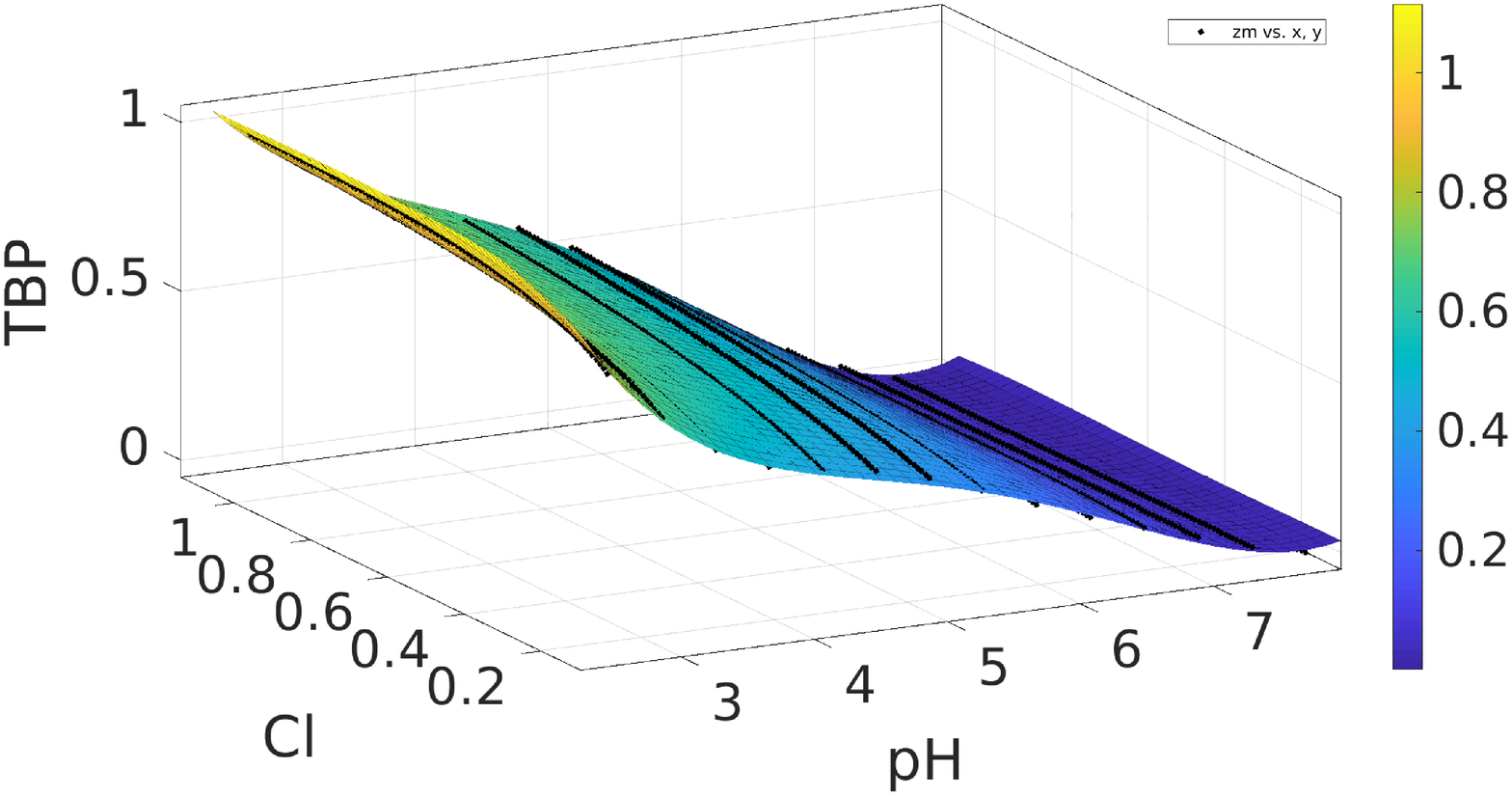}
		\caption{TBP vs pH and Cl for temperature of T=39$^o$C.  }  \label{fig:xray4a}
\end{figure}

\begin{figure}[h]
	\centering
	\includegraphics[width=10cm]{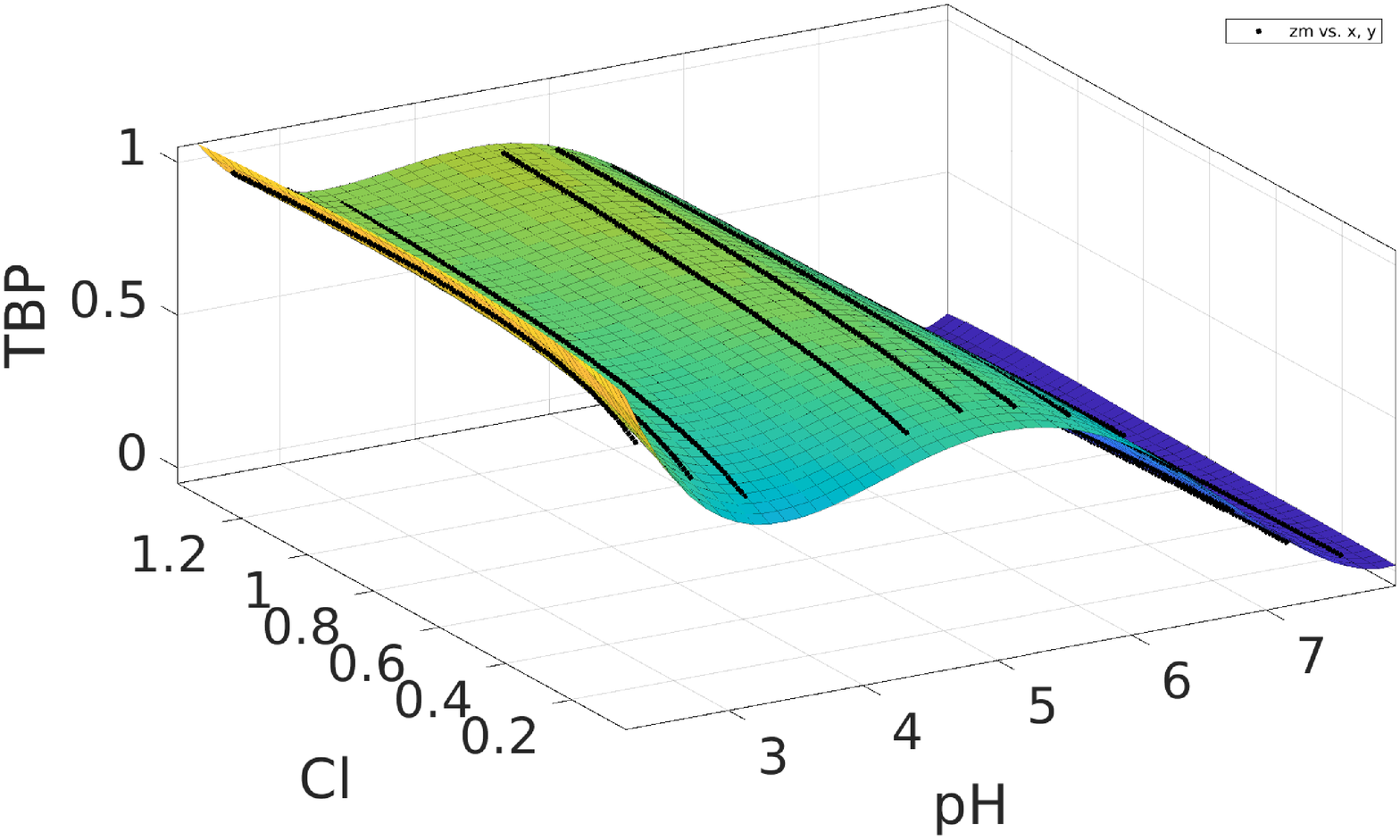}
		\caption{TBP vs pH and Cl for temperature of T=39$^o$C. }  \label{fig:xray4a2}
\end{figure}

Figures \ref{fig:xray4a} and \ref{fig:xray4a2} shows TBP for oil type 1 and 2 respectively (see Table \ref{tab1}).
In both cases, the TBP takes higher values for higher chloride concentration, which corresponds to what was reported in \cite{sari2019low}. Their results indicate that a wettability shift from oil-wet to water-wet occurs when salinity is reduced. 

\begin{figure}[h]
	\centering
	\includegraphics[width=10cm]{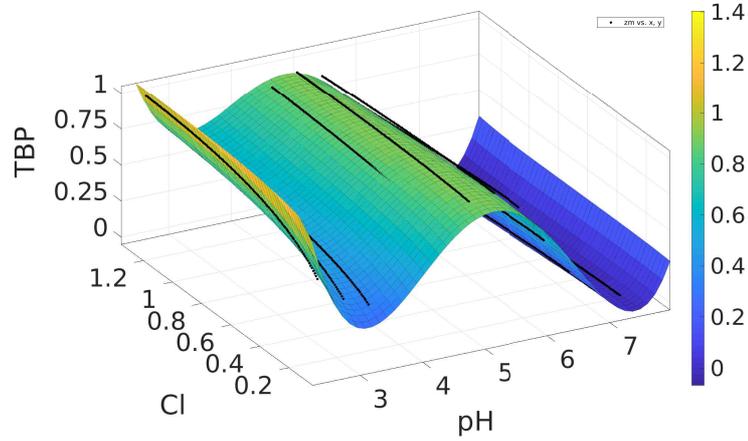}
		\caption{TBP vs pH and Cl for temperature of T=95$^o$C } \label{fig:xray4a3}
\end{figure}

The relation between TBP and $pH$ is slightly different with respect to salinity ion concentration. Figure \ref{fig:xray4a2} shows that TBP has an oscillation with respect to $pH$, i.e.,
for $pH=2.7$ a peak is observed in TBP, to later decrease and obtain a minimum at $pH=3$. Subsequently, TBP increases until it reaches a peak above $pH=5$, and then decreases for higher pH values. This behavior is maintained for temperatures of 95$^o$C but the peaks are more accentuated, see Figure \ref{fig:xray4a3}. These observed patterns of the TBP relationship with respect to pH are somewhat different from those reported by \cite{sari2019low}, where a wettability shift from water-wet to oil-wet can be achieved with increasing brine pH. 

Our numerical experiment and the experimental result from \cite{sari2019low} suggest that the relationship between TBP and wettability is not proportional to the quantity $pH$.

\begin{align}
\nonumber
   & TBP(x,y)=0.0054x^4 + 0.0091x^3y - 0.1177x^3 - 0.0041x^2y^2 - 0.1433x^2y \\
\nonumber
& + 0.9304x^2 - 0.0469xy^3 + 0.1905xy^2 + 0.5342xy - 3.269x - 0.099y^4 \\
&+ 0.615y^3   - 1.422y^2 + 0.2537y + 4.72.
\label{ed12}
    \end{align}

\begin{figure}[h]
	\centering
	\includegraphics[width=8cm]{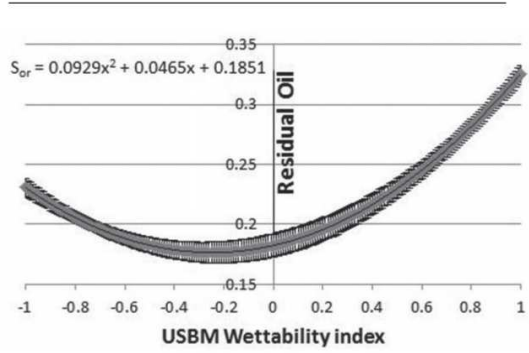}
		\caption{Residual Oil Saturation as a function of the USBM (United States Bureau of Mines) wetting
index. \cite{ebeltoft2014parameter}}  \label{fig:xray4ab}
\end{figure}

Once wettability is estimated we use the procedure proposed in \cite{bruining2021upscaling} to estimate the residual oil $S_{or}$. Given the wettability index we obtain $S_{or}$ from Figure \ref{fig:xray4ab}. Next, we obtain the initial water $S_{wi}$ using
the empirical relationship (\cite{lomeland2012versatile})
\begin{equation}
    S_{or}=2.0698S_{wi}^3-4.3857S_{wi}^2+2.17419S_{wi}+0.148.
    \label{esor}
\end{equation}
For estimating the residual oil $S_{or}$, we use the linear relationship with salt concentration given in \cite{jerauld2008modeling}.

Here, we propose a new formula for the parameter $\theta$ in equation \eqref{thetae} that takes into account the relationship between brine and pH as well as the wettability. The formula provides a direct method for calculating the parameter $\theta$,  allowing to consider the concentration of complexes in the fractional flow. The use of TBP in this context is a novel approach, as previous works have used total ionic strength instead (see e.g., \cite{korrani2016mechanistic}).

Formula for the parameter $\theta$ in \eqref{thetae} in term of TBP is given by
\begin{equation}
    \theta=\frac{TBP(pH,Cl)-TBP_l}{TBP_h-TBP_l},
    \label{thetae1a}
\end{equation}
where TBP is given in \eqref{ed12}, $TBP_l$ and $TBP_h$ represent the TBP for the lower and higher brine concentration for a fixed $pH$. For example for Oil 1 in Table \ref{tab1} and $pH=4$ we have
$TBP_l=0.3841$ and $TBP_h=0.534$. 
This formulation takes advantage of the established relationship between TBP and wettability. Therefore, once TBP is scaled, it serves to ensure the effect of salinity on the fractional flow function in an appropriate way.

\section{\textbf{Solving the Riemann-Goursat problem  by MOC}}

Once we have determined the coefficients of the system of conservation laws \eqref{massabalancecarbon1a1}-\eqref{massabalancecarbon1a14} and fixed the fractional flow function $f_w$ in \eqref{eq1a} with $\theta$ given in \eqref{thetae1a}, we can find the solution of such a system.
Notice that the fractional flow has been determined taking into account the wettability values (through TBP), which in turn are estimated from the complexes.
This solution of the system will allow us to determine the influence of the reduction of salt concentration at the point of injection.

The basis for MOC for the Riemann-Goursat problem is to assume that the 
independent variables $(s_w,pH,Cl,u)$ are functions of the variable $\xi=x/t$, which is possible
because the Riemann solution are scale invariant under the map $(t,x) \rightarrow (\alpha t, \alpha x)$; 
then we can take $U=(s_w,pH,Cl,u)(x,t)=(s_w,pH,Cl,u)(x(\xi),t(\xi))=(s_w,y,u)(\xi)$ with characteristic lines $\xi=x/t$. If we assume that the functions $U(\xi)=(s_w(\xi),y(\xi),u(\xi))$ are differentiable along the characteristic lines, the following generalized eigenvalue equation for characteristics values holds
\begin{equation}
Ar=\lambda Br, \quad \text{where} \quad A={\partial F}/{\partial U}, \quad B={\partial G}/{\partial U}.
\label{ge1}
\end{equation}
The eigenvector $r$ is parallel to $d{U}/d\xi$, so the rarefaction curves are tangent to the characteristic field given by the normalized eigenvector $r$.

The physical model consists of four unknowns states variables $s_w,pH,[Cl],u$ satisfying system \eqref{massabalancecarbon1a1}-\eqref{massabalancecarbon1a14}.
The system has four equations with the unknown variables $ (s, y, u) $
with $ y=(y_1,y_2)=(pH,[Cl]) $ given by
\begin{align}
&\frac{\partial }{\partial t}(\varphi \rho_{wj}(y)s_w+\varphi \rho_{oj}(y)s_o+(1-\varphi)\rho_{rj}(y))+
\frac{\partial }{\partial x}((u)(\rho_{wj}(y)f_w+\rho_{oj}(y)f_o)=0.\label{eq1}
\end{align}
We take $s=s_w$, $f=f_w$, $s_o=1-s$ and $f_o=1-f_{w}$.
The accumulation $G$ and flux functions $\hat{F}=(u/\phi)F$ are given by
\begin{align}
&G_j=\varphi\rho_{wj}(y)s_w+\varphi \rho_{oj}(y)s_o+(1-\varphi)\rho_{rj}(y)\\
&F_j=\rho_{wj}(y)f_w+\rho_{oj}(y)f_o
\label{acful}
\end{align}
The index $w$ (water) is often replace by the index $a$ (aqueous phase), and the index $j=1,\cdots,4$ is used to denote chemical species.

The main features in the $1$-D Riemann 
solutions of hyperbolic systems are 
the rarefaction and shock curves.
The rarefactions are obtained from integral
curves of the line fields, given by the eigenvectors
of system \eqref{ge1}, where the matrices $A,B$ represent the Jacobian matrices $A=D\hat{F}$ and $B=DG$ for the flux $\hat{F}$ and the accumulation $G$. For system $(\ref{eq1})$, the matrices $B$, $A$ are given by
\begin{align}
&B_{i,1}=[\rho_i], \quad B_{i,k+1}=\frac{\partial \rho_{wi}}{\partial y_k}s+\frac{\partial \rho_{oi}}{\partial y_k}s_o+\frac{\partial \rho_{ri}}{\partial y_k},  \text{ and } B_{i,n+1}=0.\label{matrixB} \\
&A_{i,1}= (u/\phi)[\rho_i]\frac{\partial f}{\partial s}, \quad 
A_{i,k+1}= (u/\phi)\left(\frac{\partial \rho_{wi}}{\partial y_k}f
+\frac{\partial \rho_{oi}}{\partial y_k}f_o \right)\text{ and }
A_{i,n+1}= F_i, \label{matrixA}
\end{align}
in which $i=1,\cdots,4$, $k=1,2$ and
\begin{equation}
[\rho_i]=\rho_{wi}-\rho_{oi}. \label{diferenca}
\end{equation}

From the Jacobian matrices, we obtain the eigenpairs which we have summarized in 
\begin{proposition}
	\label{lemakt}
	The eigenpairs of the eigenvalue problem $A\vec{r}=\lambda B\vec{r}$, where the matrices $B$ and $A$ represent the Jacobian
	of the accumulation and flux terms of system $(\ref{eq1})$  are the Buckley-Leverett eigenpairs ($\lambda_s,\vec{r}_s$)  given by
	\begin{equation}
	\lambda_s=\frac{u}{\varphi}\frac{\partial f}{\partial s_w}
	\quad \text{ and } \quad \vec{r}_s=(1,0,0,0)^T,\label{lambdasa}
	\end{equation}
	and two composition chemical with eigenpairs  given by $(\lambda_{\Lambda},\vec{r}_{\Lambda})$:
	\begin{equation}
	\lambda_{\Lambda}=\frac{u}{\varphi}\frac{f-\Lambda}{s-\Lambda}.\label{eig2a}
	\end{equation}
	We obtain $\Lambda$ and $\vec{v}=(v_1,v_{2})^{T}$ as the solutions of the compositional generalized eigenvalue problem 
	\begin{equation}
	((\mathcal{A}+\mathcal{A}_1)-\Lambda (\mathcal{B}+\mathcal{B}_1))\vec{v}=0,
	\label{eigeqs2}
	\end{equation}
	where the matrices $\mathcal{A}, \mathcal{A}_1, \mathcal{B}$ and  $\mathcal{B}_1$  are given in  \eqref{EB} and
	\eqref{EB1}.

    Moreover, we obtain $\vec{r}_{\Lambda}= (r_1,\cdots,r_{4})$ as 
    \begin{align}
     \label{eiget1a}
    &r_{4}=-\frac{\displaystyle \sum_{2}^{3}\mathcal{G}^{(4)}_{4,j}v_{j-1}}{\nu_{4}},~~
    r_1=\frac{\mathcal{G}_{1,4}r_{4}-\displaystyle \sum_{2}^{3}\mathcal{G}_{1,j}v_{j-1}}{\mathcal{G}_{1,1}},\\	
    &r_{j}=v_{j-1}, \quad \text{for} \quad j=2,3.
    \label{eiget1}
    \end{align}
    where the matrices $\mathcal{G}$ and $\mathcal{G}^{(4)}_{4,j}$ evaluated at $\lambda=\lambda_{\Lambda}$ are given in \eqref{matrixnew} and  \eqref{g2ka1A1}-\eqref{g2ka}.

\end{proposition}
\textbf{Proof of Proposition}

The idea of the proof consist of reducing the matrix $\mathcal{G}$ by continuous application
of Gauss procedure leaving the matrix as a reduced matrix where the eigenvalues and eigenvectors
are obtained easily.

To obtain the eigenvalues we solve $det(A-\lambda B)=0$, where $A-\lambda B=\mathcal{G}$, and $\mathcal{G}=(\mathcal{G}_{i,j})$ for $i,j=1,\cdots,4$.
Here we use an auxiliary variable $k$ that ranges from $1$ to $2$ and we write $\mathcal{G}_{i,j}$ as:
\begin{equation}
\mathcal{G}_{i,1}=[\rho_i]\xi_1, \quad
\mathcal{G}_{i,k+1}=\frac{\partial \rho_{wi}}{\partial y_k}\xi_2+
\frac{\partial \rho_{oi}}{\partial y_k}\xi_3-\lambda \frac{\partial  \rho_{ri}}{\partial y_k}
\quad   \text{ and } \quad
\mathcal{G}_{i,4}= F_i, \label{matrixnew}
\end{equation}
where we define the auxiliary variables 
\begin{equation}
\xi_1=\left(u\frac{\partial f}{\partial s}-\lambda \right), \quad
\xi_2=(u f-\lambda s)\quad \text{ and }\quad
\xi_3=(uf_o-\lambda s_o).\label{xis}
\end{equation}
For calculations purposes, that we do in the next proofs, it is useful to define the following functions $\gamma_{ij}$, 
$\varrho_{ij}$, $\nu_i$, $\vartheta_{ij}$ and $\varsigma_{ij}$  
for $i=1,\cdots,4$, $j=1,2$ as
\begin{align}
&\gamma_{ij}=\frac{\partial \rho_{wi}}{\partial y_j}[\rho_1]-
\frac{\partial \rho_{w1}}{\partial y_j}[\rho_i],
\quad
\varrho_{ij}=\frac{\partial \rho_{oi}}{\partial y_j}[\rho_1]-
\frac{\partial \rho_{o1}}{\partial y_j}[\rho_i],
\label{coef1}
\end{align}
\begin{align}
&\pi_{ij}=\frac{\partial \rho_{ri}}{\partial y_j}[\rho_1]-
\frac{\partial \rho_{r1}}{\partial y_j}[\rho_i],
\quad 
\nu_i=[\rho_1]F_{i}-[\rho_i]F_{1},
\label{coef2}
\end{align}
\begin{align}
\label{gamma1}
 &\vartheta_{ij}=\gamma_{ij}\nu_{4}-\gamma_{4,j}\nu_i, \quad
 \varsigma_{ij}=\varrho_{ij}\nu_{4}-\varrho_{4,j}\nu_i,\\
&  \tau_{ij}=\pi_{ij}\nu_{4}-\pi_{4,j}\nu_i.
 \label{gamma}
 \end{align}
All these coefficients (\eqref{coef1}- \eqref{gamma}) depend only of the variables $y$. Now, we are able to perform the proof os results in the text. 

\begin{itemize}
	
\item[1)] Substituting  the $i$-th row of matrix $(\ref{matrixnew})$, for
$i=2,\cdots,4$,  by the sum of the first row of $(A-\lambda B)$ (the elements  of which are given  by (\ref{matrixnew}))
multiplied by $-[\rho_i]$ with its $i$-th row   multiplied by 
$[\rho_1]$, we obtain, for $j=1,2$ and $i=2,3,4$
\begin{equation}
\mathcal{G}\sim\begin{pmatrix}
\mathcal{G}_{1,1} & \mathcal{G}_{1,j+1} &\mathcal{G}_{1,4}\\
\mathbb{O} & \mathcal{G}^{(1)}_{i,j+1} & \mathcal{G}^{(1)}_{i,4}\\
\end{pmatrix}.\label{matrix3}
\end{equation}
Here $\mathbb{O}$ is the column vector of three zeros and block matrices $(\mathcal{G}^{(1)}_{i,j+1})$ and $ (\mathcal{G}^{(1)}_{i,4})$ for
$i=2,3,4$ and $j=1,2$ are the block matrices whose elements are given by
\begin{equation}
\mathcal{G}^{(1)}_{i,j+1}=\left(\gamma_{ij}\xi_2+\varrho_{ij}\xi_3-\pi_{ij}\lambda\right)_{1\leq j\leq 2}\quad 
\text{ and }
\quad 
\mathcal{G}^{(1)}_{i,4}=\nu_{i},
\end{equation}
where $\gamma_{ij}$, $\varrho_{ij}$ $\pi_{ij}$ and  $\nu_{i}$ are given by Eq. \eqref{coef1} and \eqref{coef2}.

Notice here that if $[\rho_i]$, given y $(\ref{diferenca})$, is zero for some index $i$, 
the corresponding position in the first column is zero and we do not need to perform
calculations to vanish this position. Moreover, if $[\rho_1]$ is zero, we exchange the position
of row $1$ with another row to obtain a non-zero pivot. 

\item[2)] Now, we substitute  $i$-th row of $(\ref{matrix3})$, for $i=2,3$,  by the sum of the four row of $(\ref{matrix3})$
multiplied by $-\nu_i$ with the $i$-th row of $(\ref{matrix3})$ multiplied by $\nu_{4}$, and we obtain for $j=1,2$ and
$i=2,3$
\begin{equation}
\mathcal{G}\sim\begin{pmatrix}
\mathcal{G}_{1,1} & \mathcal{G}_{1,j+1} &\mathcal{G}_{1,4}\\
\mathbb{O} & \mathcal{G}^{(2)}_{i,j+1} &\mathbb{O}\\
0 & \mathcal{G}^{(2)}_{4,j+1}&\nu_{4}\\
\end{pmatrix}.\label{pms}
\end{equation}
Here $\mathbb{O}$ is the column vector of two zeros and $(\mathcal{G}^{(2)}_{i,j+1})$ and $ (\mathcal{G}^{(1)}_{4,j+1})$ for
$i=2,3$ and $j=1,2$ are the block matrices given by
\begin{equation}
\mathcal{G}^{(2)}_{i,j+1}=\vartheta_{ij}\xi_2+\varsigma_{ij}\xi_3-\tau_{ij}\lambda
,~
\mathcal{G}^{(2)}_{4,j+1}=\left(\gamma_{4j}\xi_2+\varrho_{4j}\xi_3-\pi_{4j}\lambda\right), \label{g2k}
\end{equation}
where $\vartheta_{ij}$, $\varsigma_{ij}$ and $\tau_{ij}$ are given by \eqref{gamma1} and $(\ref{gamma})$, for $i=1,\cdots,4$, $j=1,2$. 

\item[3)] Now, we substitute the $i$-th row of $(\ref{pms})$, for $i=2,3$,  by the sum of the four row of $(\ref{pms})$
multiplied by $-\tau_{i1}$ with the $i$-th row of $(\ref{pms})$ multiplied by $\pi_{41}$, and we obtain for $j=1,2$ and
$i=2,3$
\begin{equation}
\mathcal{G}\sim\begin{pmatrix}
\mathcal{G}_{1,1} & \mathcal{G}_{1,j+1} &\mathcal{G}_{1,4}\\
\mathbb{O} & \mathcal{G}^{(3)}_{i,j+1} &\mathbb{O}\\
0 & \mathcal{G}^{(3)}_{4,j+1}&\nu_{4}\\
\end{pmatrix}.\label{pmsa1}
\end{equation}
Here $\mathbb{O}$ is the column vector of two zeros and $(\mathcal{G}^{(3)}_{i,j+1})$ and $ (\mathcal{G}^{(3)}_{4,j+1})$ for
$i=2,3$ and $j=1,2$ are the block matrices given by
\begin{equation}
\mathcal{G}^{(3)}_{i,2}=m_{i1}\xi_2+n_{i1}\xi_3,~\mathcal{G}^{(3)}_{i,3}=m_{i2}\xi_2+n_{i2}\xi_3-\alpha_{i2}\lambda
\label{g2ka1a}
\end{equation}
\begin{equation}
\mathcal{G}^{(3)}_{4,j+1}=\left(\gamma_{4j}\xi_2+\varrho_{4j}\xi_3-\pi_{4j}\lambda\right), \label{g2ka12}
\end{equation}
where
\begin{align}
&m_{ij}=\vartheta_{ij}\pi_{41}-\gamma_{4j}\tau_{i1},~n_{ij}=\varsigma_{ij}\pi_{41}-\rho_{4j}\tau_{i1},\\
&\alpha_{i2}=\tau_{i2}\pi_{41}-\pi_{42}\tau_{i1},~\alpha_{i1}=0.
\label{coef1ma12}
\end{align}
and $\vartheta_{ij}$, $\varsigma_{ij}$ and $\tau_{ij}$ are given by \eqref{coef2} and $(\ref{gamma})$, for $i=1,\cdots,4$, $j=1,2$. 

\item[4)]
Now, we substitute  third row of $(\ref{pmsa1})$, by the sum of the second row of $(\ref{pmsa1})$
multiplied by $-\alpha_{32}$ with the third row of $(\ref{pmsa1})$ multiplied by $\alpha_{21}$, and we obtain 
\begin{equation}
\mathcal{G}\sim\begin{pmatrix}
\mathcal{G}_{1,1} & \mathcal{G}_{1,j+1} &\mathcal{G}_{1,4}\\
\mathbb{O} & \mathcal{G}^{(4)}_{i,j+1} &\mathbb{O}\\
0 & \mathcal{G}^{(4)}_{4,j+1}&\nu_{4}\\
\end{pmatrix}.\label{pmsa2}
\end{equation}
Here $\mathbb{O}$ is the column vector of two zeros and $(\mathcal{G}^{(4)}_{i,j+1})$ and $ (\mathcal{G}^{(4)}_{4,j+1})$ for
$i=2,3$ and $j=1,2$ are the block matrices given by:
\begin{equation}
\mathcal{G}^{(4)}_{2,2}=\bar{m}_{21}\xi_2+\bar{n}_{21}\xi_3,~\mathcal{G}^{(4)}_{2,3}=\bar{m}_{22}\xi_2+\bar{n}_{22}\xi_3+\alpha_{22}\lambda
\label{g2ka1A1}
\end{equation}
\begin{equation}
\mathcal{G}^{(4)}_{3,2}=\bar{m}_{31}\xi_2+\bar{n}_{31}\xi_3,~\mathcal{G}^{(4)}_{3,3}=\bar{m}_{32}\xi_2+\bar{n}_{32}\xi_3
\label{g2ka1}
\end{equation}
\begin{equation}
\mathcal{G}^{(4)}_{4,j+1}=\left(\gamma_{4j}\xi_2+\varrho_{4j}\xi_3+\pi_{4j}\lambda\right), \label{g2ka}
\end{equation}
where
\begin{equation}
\bar{m}_{21}=m_{21}, \bar{n}_{21}=n_{21}, \bar{m}_{22}=m_{22} , \bar{n}_{22}=n_{22},
\end{equation}
and
\begin{align}
\bar{m}_{31}=m_{31}\alpha_{21}-m_{21}\alpha_{32},~\bar{n}_{31}=n_{31}\alpha_{21}-n_{21}\alpha_{32}.
\label{coef1ma1}
\end{align}
\begin{align}
\bar{m}_{32}=m_{32}\alpha_{21}-m_{22}\alpha_{32},~\bar{n}_{32}=n_{32}\alpha_{21}-n_{22}\alpha_{32}.
\label{coef1ma1a}
\end{align}


\item[5)]
Since $\mathcal{G}^{(4)}_{i,j+1}$ is block matrix, but this matrix appears in the rows $2$ to $3$ and columns $2$ to $3$ it is useful to define 
matrix $\left(G^{(4)}_{l,r}\right)$ for $l,r=1,\cdots,2$  from $\left(\mathcal{G}^{(4)}_{i,j+1}\right)$
\begin{equation}
G^{(4)}_{l,r}=\mathcal{G}^{(4)}_{l+1,r+1} \quad \text{ for } l,r=1,\cdots,2.\label{gge}
\end{equation}

From $(\ref{pmsa2})$ and $\det(A-\lambda B)=0$ 
a solution is $\xi_1=0$. Since $\xi_1$ is given by $(\ref{xis}.a)$, 
thus we obtain the eigenpair $(\lambda_s,\vec{r}_s)$  given 
by $(\ref{lambdasa})$. 
For this eigenpair, only saturation changes and we identify this family wave as \textit{saturation wave}, or
Buckley-Leverett type wave. 

Using $(\ref{pmsa2})$ and $det(A-\lambda B)=0$, the other eigenvalues are obtained by solving
\begin{equation}
\det({G}^{(4)}_{l,r})=0 \quad \text{ for } \quad l,r= 1,2,\label{eigeqs}
\end{equation}
with ${G}^{(4)}_{l,r}$ given by \eqref{g2ka1A1} and \eqref{g2ka1}. 

To obtain the corresponding eigenvalue, we substitute $\lambda_\Lambda$ 
given by $(\ref{eig2a})$ into $(\ref{eigeqs})$ and using that 
\begin{equation}
\xi_2=u f-\lambda_\Lambda s= u\frac{f(s-\Lambda)-(f-\Lambda)s}{s-\Lambda}=u\Lambda\frac{s-f}{s-\Lambda}.\label{ueq1}
\end{equation}
\begin{equation}
\xi_3=uf_o-\lambda_\Lambda s_o=u(1-f)-\lambda_\Lambda(1-s)=u(1-\Lambda)\frac{s-f}{s-\Lambda}.\label{ueq1n}
\end{equation}
Substituting $\xi_2$ and $\xi_3$ in \eqref{g2ka1A1} and \eqref{g2ka1} the matrix in \eqref{eigeqs} can be rewrite as 
\begin{equation}
\left(\begin{array}{cc} \bar{n}_{21}+\Lambda(\,m_{21}-\,\bar{n}_{21}) & \bar{n}_{22}+\Lambda(\,m_{22}-\,\bar{n}_{22})\\ \bar{n}_{31}+\Lambda(\,m_{31}-\,\bar{n}_{31}) & \bar{n}_{32}+\Lambda(\,m_{32}-\,\bar{n}_{32}) \end{array}\right)+\left(\begin{array}{cc} 0 & \frac{\alpha_{22}(f-\Lambda)}{(s-f)}\\ 0 & 0 \end{array}\right)
\label{mm1}
\end{equation}
Denoting by 
\begin{equation}
\mathcal{A}=\left(\begin{array}{cc} \bar{n}_{21} & \bar{n}_{22}\\ \bar{n}_{31} & \bar{n}_{32} \end{array}\right)
~~~~\text{and}~~~
\mathcal{B}=-\left(\begin{array}{cc} \,m_{21}-\,\bar{n}_{21} & \,m_{22}-\,\bar{n}_{22}\\ \,m_{31}-\,\bar{n}_{31} & \,m_{32}-\,\bar{n}_{32} \end{array}\right),
\label{EB}
\end{equation}
we obtain
\begin{equation}
\mathcal{A}-\Lambda\mathcal{B}+\left(\begin{array}{cc} 0 & \frac{\alpha_{22}(f-\Lambda)}{(s-f)}\\ 0 & 0 \end{array}\right).
\label{mm12}
\end{equation}
Also, denoting by
\begin{equation}
\mathcal{A}_1=\left(\begin{array}{cc} 0 & \frac{\alpha_{22}f}{(s-f)}\\ 0 & 0 \end{array}\right)~~~~\text{and}~~~
\mathcal{B}_2=\left(\begin{array}{cc} 0 & \frac{\alpha_{22}}{(s-f)}\\ 0 & 0 \end{array}\right),
\label{EB1}
\end{equation}
we have
\begin{equation}
{G}^{(4)}_{l,r}=(\mathcal{A}-\Lambda\mathcal{B})+(\mathcal{A}_1-\Lambda\mathcal{B}_2),
\end{equation}
where the matrix $\mathcal{A}$ and $\mathcal{B}$ only depend the variables $y$, while  the matrix $\mathcal{A}_1$ and $\mathcal{B}_1$ depend on $y$ and $s$.

In this notation is easy to verify that equation \eqref{eigeqs} is
a quadratic equation in the variable $\Lambda$. Thus equation \eqref{eigeqs2}  is a $(2)\times (2)$ algebraic system. If we have two different roots, we obtain two more eigenvalues of form
$(\ref{eig2a})$.  We call each eigenvalues as $\lambda_{\Lambda_i}$ for $i=1,2$.

Also, notice that if $\rho_{rj}=0$ $(j=1,\cdots,4)$ then $\alpha_{22}=0$
and the co\-rres\-pon\-ding roots of equation \eqref{eigeqs} only depend on the chemical variable $y$. In this case solution can be found more easily (see \cite{alvarez2023nonlinear1}).

\item[6)] The eigenvectors related with $\lambda_{\Lambda}$ are obtained solving
\begin{equation}
\begin{pmatrix}
\mathcal{G}_{1,1} & \mathcal{G}_{1,j+1} &\mathcal{G}_{1,4}\\
\mathbb{O} & \mathcal{G}^{(4)}_{i,j+1} &\mathbb{O}\\
0 & \mathcal{G}^{(4)}_{4,j+1}&\nu_{4}\\
\end{pmatrix}\vec{r}=0.\label{pmsno}
\end{equation}
We can split the calculation of $\vec{r}$. 
First we obtain 
the coordinates $(r_2,r_3)$ of eigenvector 
$\vec{r}_\Lambda=(r_1,r_2,r_3,r_{4})$. 
To do our calculations, we define the auxiliary vector $\vec{v}$
of two coordinates, which is the solution of
\begin{equation}
\left({G}^{(4)}_{l,r}\right)\vec{v}^T=0\quad \text{ for } \quad l,r=1,2.\label{pmsnok}
\end{equation}
Using  $(\ref{ueq1})$-$(\ref{ueq1n})$, we can see that
$(\ref{pmsnok})$, after simplifications is written again as $(\ref{eigeqs2})$.

The coordinates $r_{n+1}$ and $r_1$ are  obtained
by solving the first and the last equations of $(\ref{pmsno})$ and using 
$(\ref{ueq1})$-$(\ref{ueq1n})$, after some tedious
calculations, we obtain Eqs. $(\ref{eiget1a})-(\ref{eiget1})$.  $\square$ 
\end{itemize}
The integral curves $\mathcal{W}^s$  and $\mathcal{W}^{\Lambda}$ associated to $\vec{r}_s$ and $\vec{r}_{\Lambda}$ are obtained by integrating the ODEs  
\begin{equation}
d(s,y,u)/d \xi=\vec{r}_s~\text{and}~\frac{d(s,y,u)}{d \xi}=\vec{r}_{\Lambda}.
\label{integral}
\end{equation}
From \eqref{integral}$a$, we obtain the Buckley-Leverett(B-L) wave, where only the saturation changes, which is denotes by $\mathcal{R}_s$. Moreover, the solution of \eqref{integral}$b$ determines the waves associated to $\lambda_{\Lambda}$ denotes by $\mathcal{R}_{\Lambda}$. Here, we do the technical proofs of results which appear in paper \cite{alvarez2020resonance,alvarez2023nonlinear1}. These papers describe the numeric and theory of
the main wave interactions of the
system of conservation laws studied here.

\subsection{Rankine-Hugoniot Locus}
\label{rsk1}

In this section we summarized how to calculated the discontinuous solution of the system of conservation laws studied here.
Details of proof can be found in \cite{alvarez2020resonance}.
We denote $y=(pH,Cl)$ to gain in clearance.

The discontinuous solution of system of conservation laws
$(\ref{eq1})$ satisfy the Rankine-Hugoniot condition, i.e. for a given left and right state
$(s^-,y^-,u^-)$ and $(s^+,y^+,u^+)$ respectively we have
\begin{equation}
u^+F_i((s^+,y^+)-u^-F_i(s^-,y^-)=v^s(G_i(s^+,y^+)-G_i(s^-,y^-)),
\label{rh1aa}
\end{equation}
with $i=1,\dots,n+1$ and  $v^s$ is the speed of discontinuity.
Equation \eqref{rh1aa} can be rewritten as
\begin{align}
\Phi_{i}.[v^s,u^+,u^-]=0, 
\label{rank1}
\end{align}
where $\Phi_{i}=(\Phi_{i1},\Phi_{i2},\Phi_{i2})$, with\\
$\Phi_{i1}=\s^+(\rho_{wi}^+-\rho_{oi}^+)+\rho_{oi}^+-(\s^-(\rho_{wi}^--\rho_{oi}^-)+\rho_{oi}^-)+(\rho_{ri}^+-\rho_{ri}^+)$,~\\
$\Phi_{i2}=-(\f^+(\rho_{wi}^+-\rho_{oi}^+)+\rho_{oi}^+)$,~\\ $\Phi_{i3}=(\f^+(\rho_{wi}^--\rho_{oi}^-)+\rho_{oi}^-)$ 
and  $\rho^+=\rho(\y^+)$, $\rho^-=\rho(\y^-)$,  $\f^+=\f(\s^+,y^+)$, $\f^-=\f(\s^-,y^-)$.

For each fixed state $(\s^-,y^-)$, the $\textit{Hugoniot-locus}$ $\mathcal{HL}(\s^-,y^-)$ consist of all states $(\s^+,y^+)$
satisfying $(\ref{rank1})$. In \cite{lambert2019nonlinear} is prove that
\begin{equation}
\mathcal{HL}(\s^-,y^-)=\left\{(\s^+,y^+): det(\Phi_{i}^T,\Phi_{k}^T,\Phi_{j}^T)=0\right\},
\label{hl1a}
\end{equation}
for all the combination of distinct index $\{i,j,k\}\in \{1,2,\cdots,n+1\}$.
Also, we verify that instead to consider all the combination it is enough to reduce $n-1$ equations. This results can summarized as follows: let $i_1$ and $i_2$ $\in\{1,2,\cdots,n+1\}$ two indices 
such that $\Phi_{i_1}$ and $\Phi_{i2}$ are independent linearly, then Equation \eqref{hl1a} reduces to
\begin{equation}
\mathcal{HL}(\s^-,y^-)=\left\{(\s^+,y^+): det(\Phi_{k}^T,\Phi_{i_1}^T,\Phi_{i_2}^T)=0\right\},
\label{hl2a}
\end{equation}
for $k \in\{1,2,\cdots,n+1\}$ distinct of  $i_1$ and $i_2$. Equation \eqref{hl2a} represents
a curve in the three dimensional space $(s,y)$ consisting in the intersection of two surfaces.

For fixed $(s^-,y^-)$, there exists a branch of $\mathcal{HL}(s^-,y^-)$ consisting of the 
	states of the form $(s,y)\in\Omega$, with $s$ variable and $y=y^-$.  This branch is called \textit{Buckley-Leverett or saturation branch}, and it is denoted by $\mathcal{H}_s$. The other branches are denoted by $\mathcal{H}_{\Lambda}$.
 
\subsection{Numerical and semi analytical solution}

In this work we find the solution for the system \eqref{massabalancecarbon1a1}-\eqref{massabalancecarbon1a14} by using the COMSOL solver and the in house Riemann solver called RPNfilho. 
We are interested in the Riemann-Goursat problem  for
these equations with piecewise constant initial data
\begin{equation}
\left\{
\begin{array}
[c]{ll}%
L=(Sw_{L},pH_L,Cl_L,u_L) & \textbf{if}\hspace{0.2cm}x<0,\\
R=(Sw_{R},pH_R,Cl_R,\;\cdot\;) &
\textbf{if}\hspace{0.2cm}x>0.
\end{array}
\right.\label{riemandata}
\end{equation}

The Riemann solution is obtained by the wave curve method. The solution is constructed by means of a sequence of elementary waves $w_k$ (shocks and  rarefactions) for $k=1,2,\cdots,m$
and constant states $U_k$ for $k=1,2,\cdots,p$, in which $p$ is not known a priori.

At any rate, this sequence of waves can be written as
\begin{equation}
\mathcal{U}_L\equiv \mathcal{U}_0 \overset{w_1}{\longrightarrow}
\mathcal{U}_1 \overset{w_2}{\longrightarrow}  \cdots
\overset{w_m}{\longrightarrow} \mathcal{U}_m\equiv \mathcal{U}_R
\label{sequence},
\end{equation}
where $\mathcal{U}=(s,pH,Cl,u)$.   In
the Riemann solution it is necessary that the waves have increasing speed, satisfying the so called \textit{geometrical
	compatibility}. Sometimes, this geometrical compatibility
is sufficient to furnish existence and uniqueness of the solution. Moreover, this condition is used to select the physical sequence of waves for the Riemann solution. We do not go into details about the wave curve method in this paper. Details can be found in \cite{alvarez2020resonance}.

The waves consist of the B-L saturation wave curve $\mathcal{R}_s$ where only the saturation varies, the shock wave curve $\mathcal{H}_s$, the locus where only the saturation varies and the chemical saturation wave curves $\mathcal{R}_{\Lambda_i}$ ($i=1,2$) associated to the couple
	$(\lambda_{\Lambda_i},r_{\Lambda_i})$. Moreover other compatible wave curves include the composite wave curve $C_s=\mathcal{R}_s \cup \mathcal{H}_s$, formed by a characteristic B-L shock curve, and the composite wave curves $C_{\Lambda_i}=\mathcal{R}_{\Lambda_i} \cup \mathcal{H}_{\Lambda_i}$, where $\mathcal{H}_{\Lambda_i}$ is a characteristic shock curve associated to the branch of the family $\lambda_{\Lambda_i}$ with $i=1,2$ and
 the constant state $\mathcal{C}$.

\section{Numerical solution}

In this section we present the numerical solution for the system \eqref{massabalancecarbon1a1}-\eqref{massabalancecarbon1a14} by using COMSOL and the Riemann solver RPNfilho. 
We solve the Riemann-Goursat problem for some particular initial (right) and injection (left) conditions  
for such system of equations. We do not impose any condition on the variable $u_R$ in \eqref{riemandata} because its value is obtained from the 
other variables together with the solution of the system.
Physically, the left state represents the value of the variables $pH$, $[Cl]$, $S_w$ and $u$ at the injection point. The right state presents the initial condition of such variables in the reservoir. The states studied here correspond to typical values of interest for petroleum engineering.

We assume that the fluid is incompressible, but there is mass transfer between phases of the carbon dioxide and the partial molar volume differs between phases, thus a
variable total Darcy velocity ensues. 

\subsection{RPNfilho and COMSOL solutions}

We are interested in evaluating
 the salt concentrations reduction at the injection point.
To do so, first we cross-validated the method and second we identify the main waves present in the simulation of relevant examples. 
Besides calculating the speeds of the water saturation front and of the saline front, we estimate the behavior of the $pH$.

In this simulation we take a reduction of 5$\%$ in the initial chloride concentration in the reservoir, which we take as 0.3 mole per kilogram of water. To the chloride concentration of 0.015 mole per kilogram of water  corresponds an oil saturation $S_{or}=0.228$ and an initial water $S_{wi}=0.0398$. The initial condition for the Riemann problem consists of  $L=(0.7604,4.0,0.015,1.0 \times 10^{-5})$ and
$R=(0.0398,4.0,0.3)$.

\begin{figure}[h]
	\centering
	\includegraphics[scale=0.13]{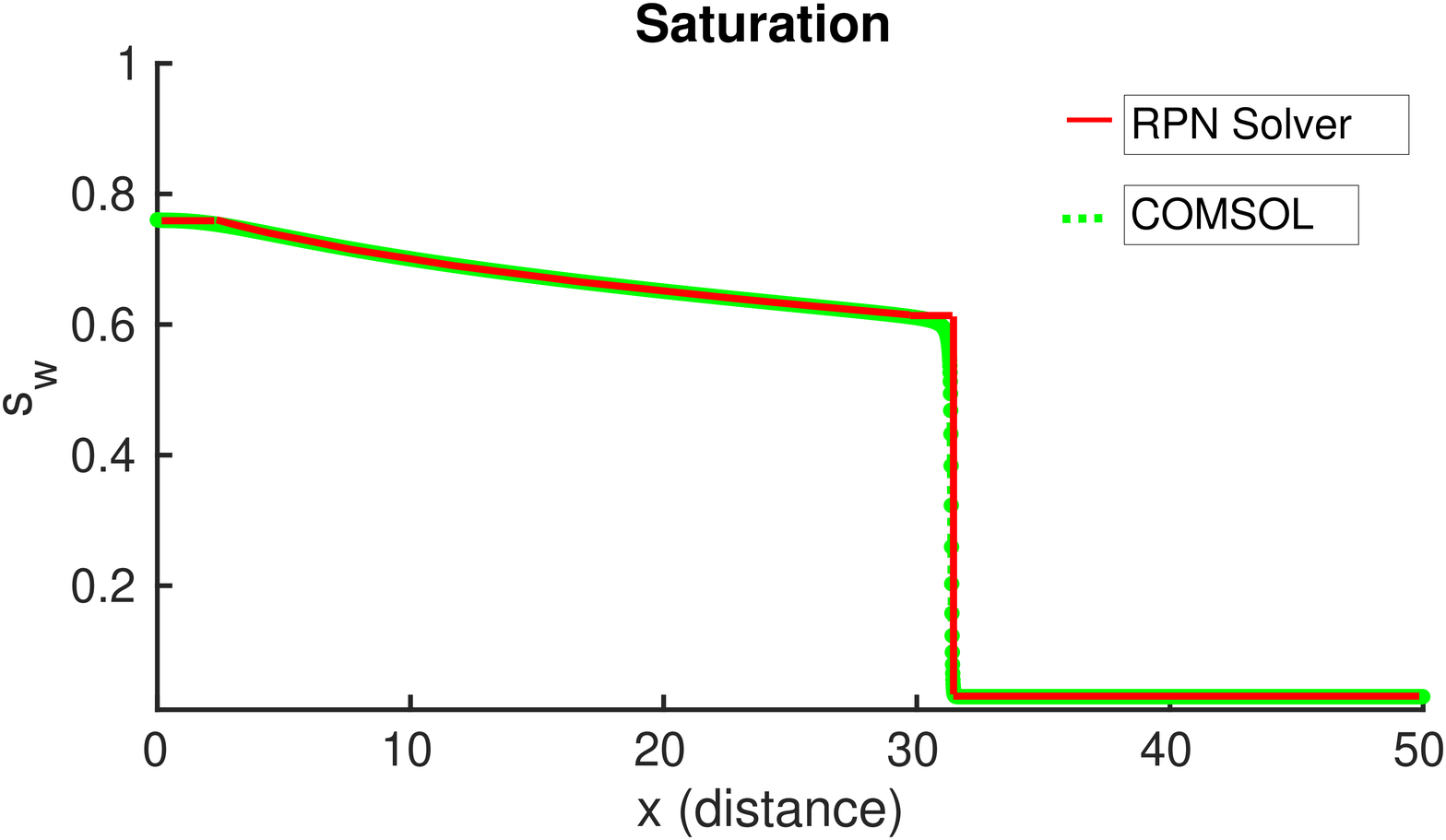}
	\includegraphics[scale=0.13]{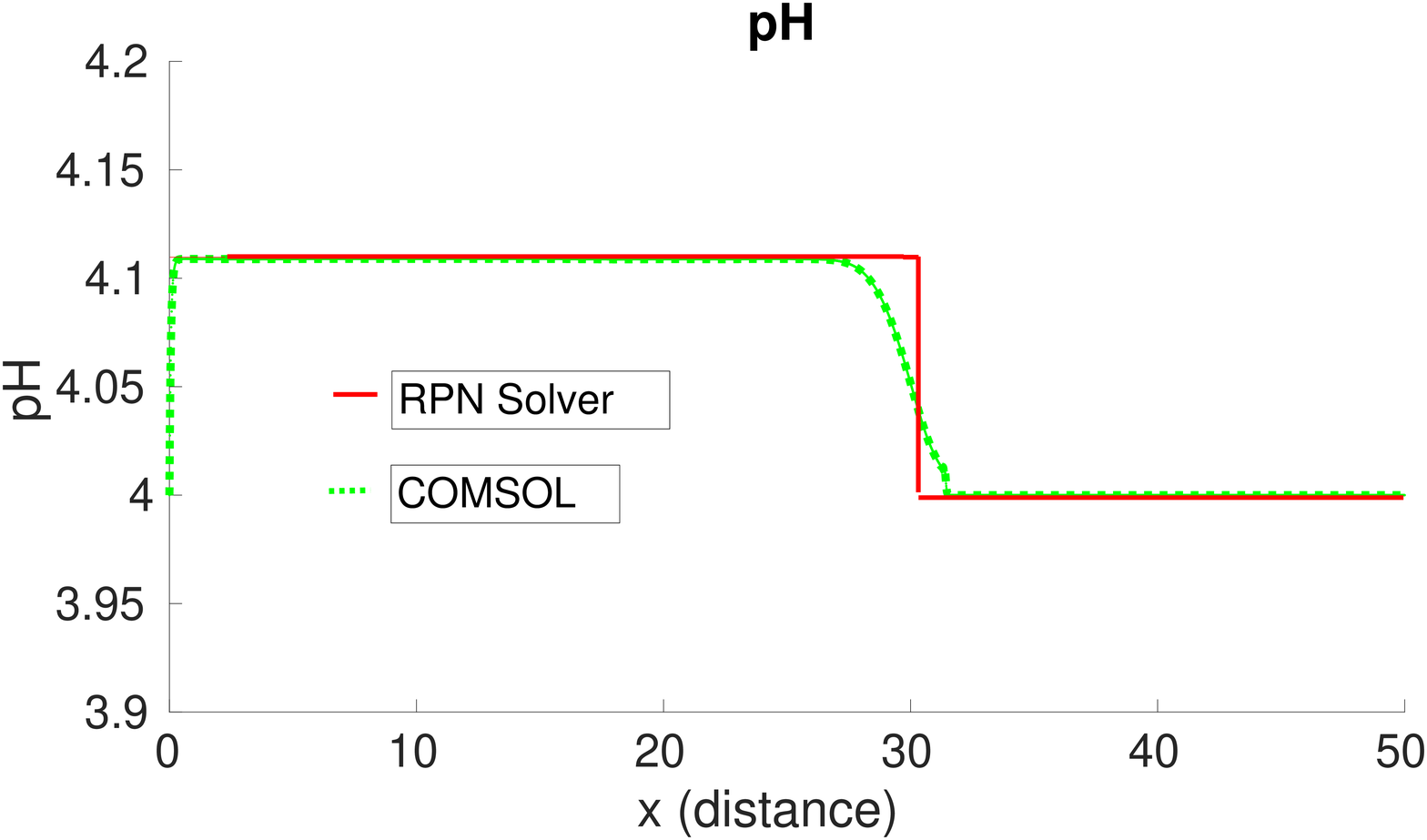}  
	\caption{a) Saturation profile from RPNfilho and COMSOL solvers, b) $pH$ profile with RPNfilho and COMSOL solvers.} 
	\label{fignova3a1}
\end{figure}

\begin{figure}[h]
	\centering
	\includegraphics[scale=0.18]{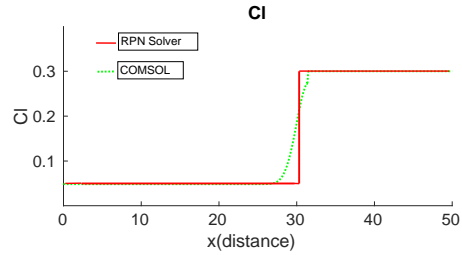}
	\caption{Chloride profile from RPNfilho and COMSOL solver} 
	\label{fignova3m}
\end{figure}

Figures \ref{fignova3a1}-\ref{fignova3m} show the water saturation, the chloride and the $pH$ profiles corresponding to the Riemann problem for a given $L$ and $R$. We verify that in this example the solution of RPNfilho and COMSOL solvers match well. The saturation profile for both models coincide. The pH and chloride profiles match well except in the shock front solution.

The structure of the Riemann solution obtained by RPNfilho solver consists of a chemical 2-rarefaction wave, a constant state, a  B-L rarefaction, a constant state, a 2-shock and a B-L shock. This sequence of waves can be represented schematically as $(L)\xrightarrow{\mathcal{R}_{\Lambda_2}}(A) \xrightarrow{\mathcal{R}_{s}}(B)\xrightarrow{\mathcal{H}_{\Lambda_2}}(C)
 \xrightarrow{\mathcal{H}_s}(R)$. This scheme means that from state $L$ to the state $A=(0.7604,4.12,0.015)$ there exists a rarefaction $\mathcal{R}_{\Lambda_2}$ of the chemical family associated to eigenvalues $\lambda_{\Lambda_2}$. This wave is followed by a constant state from the state $A$ to the state $B=(0.61,4.12,0.015)$, then we have a B-L rarefaction wave.
Then we have another constant state followed by a type 2 shock to the point $C$. Immediately afterwards a constant state follows a B-L shock and reaches the right state $R$.

In the subsequent numerical examples the structure of the solution described above does not change. In these simulations only the relative position of the intermediate states vary when the initial conditions of the rock (right state) and the injection parameters (formation water, left state) change.

\subsection{Effect of brine reduction}

In the following numerical experiments, we aim at  investigating the effect of reducing the salinity on the oil recovery. In the first case, we assume a 5$\%$ reduction in the initial salt concentration, and the injection and initial states are defined by $L=(0.76,4.0,0.015,1.0 \times 10^{-5})$ and $R=(0.0398,4.0,0.3)$, respectively. The corresponding residual oil saturation is $S_{or}=0.228$ (as reported in \cite{jerauld2008modeling}), and the initial water saturation is $S_{wi}=0.0398$ (calculated using formula \eqref{esor}).

In the second example, we assume a 20$\%$ reduction in the rock salt concentration, and the initial and injection states are defined by $R=(0.045,4.0,0.3)$ and $L=(0.76,4.0,0.06,1.0 \times 10^{-5})$, respectively. The corresponding residual oil saturation is $S_{or}=0.239$, and the initial water saturation is $S_{wi}=0.045$ (calculated using formula \eqref{esor}).

\begin{figure}[h]
	\centering
	\includegraphics[scale=0.2]{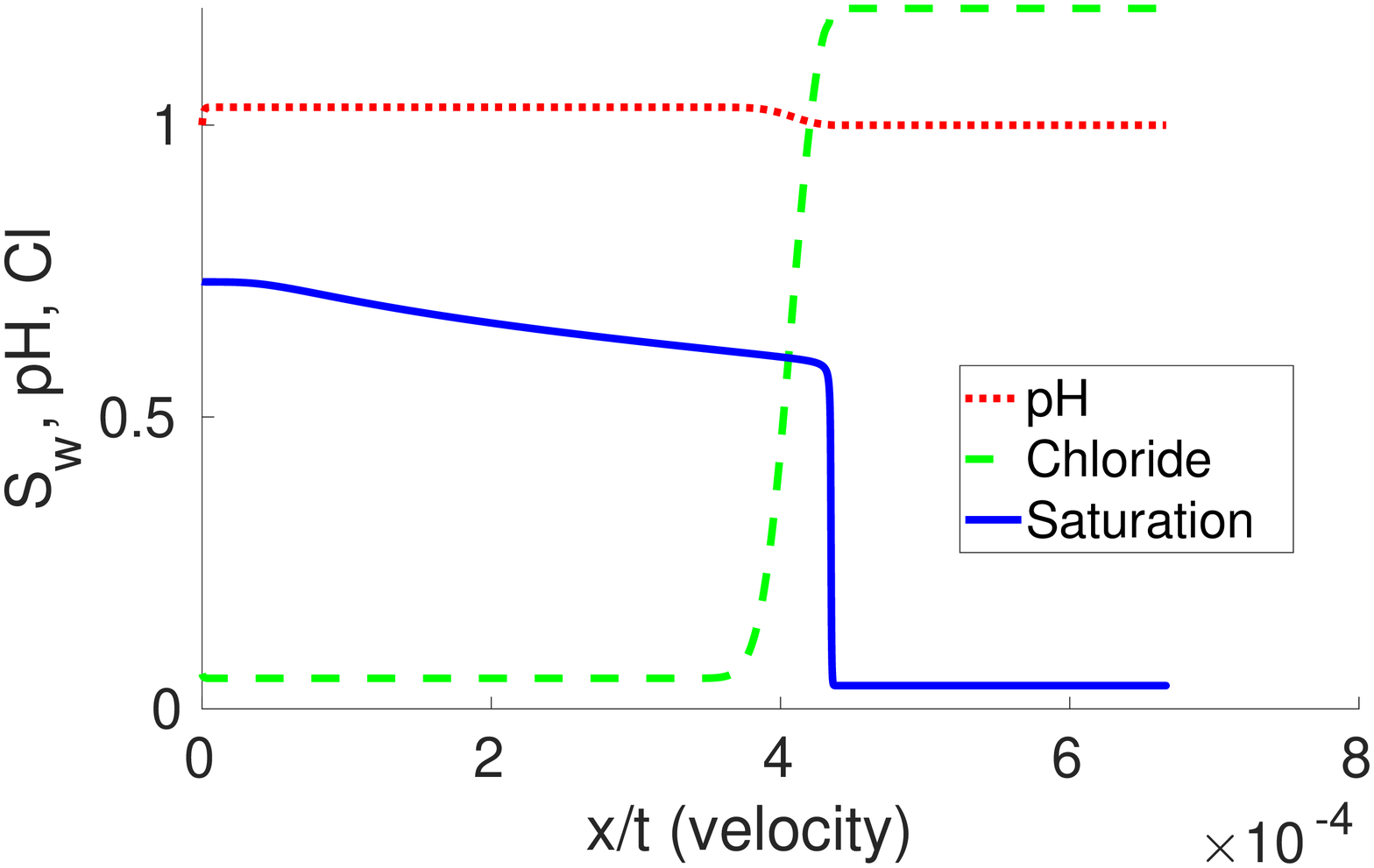}
	\includegraphics[scale=0.2]{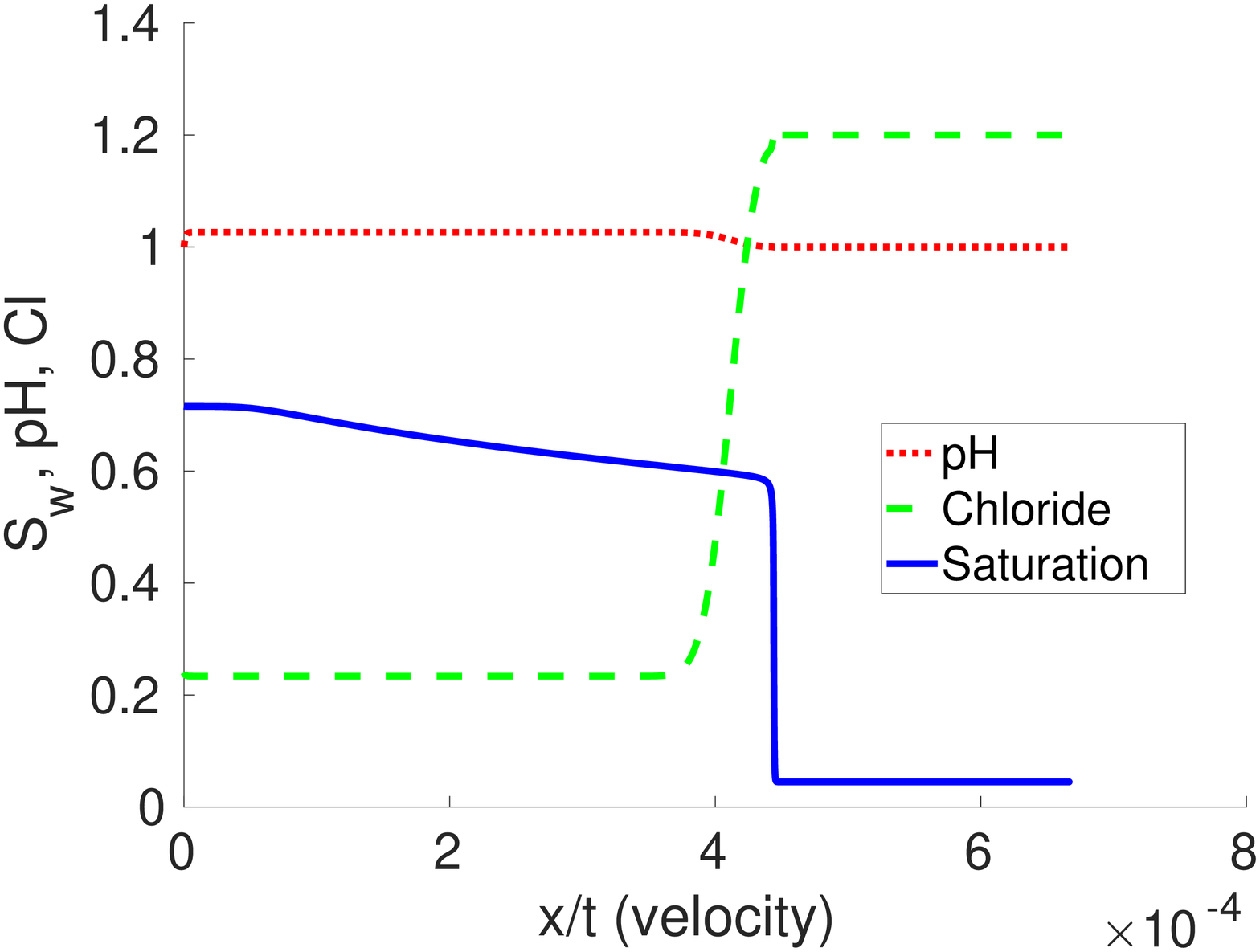}  
	\caption{The figures show profiles of the saturation $S_w$, scaled pH ($pH/4$), and scaled chloride concentration  ($[Cl]*4$) at different velocities for a 5$\%$ and 20$\%$ reduction in salt concentration (left and right figures, respectively). The velocity of the saline and pH fronts are $4.05 \times 10^{-5}$ and $4.08 \times 10^{-5}$, respectively, for the 5$\%$ reduction case, and the velocity of the saturation front is $4.34 \times 10^{-5}$. For the 20$\%$ reduction case, the velocity of the saline and pH fronts are $4.08 \times 10^{-5}$, while the velocity of the saturation front is $4.44 \times 10^{-5}$.} 
	\label{fignova3a}
\end{figure}

\begin{figure}[h]
	\centering
	\includegraphics[scale=0.17]{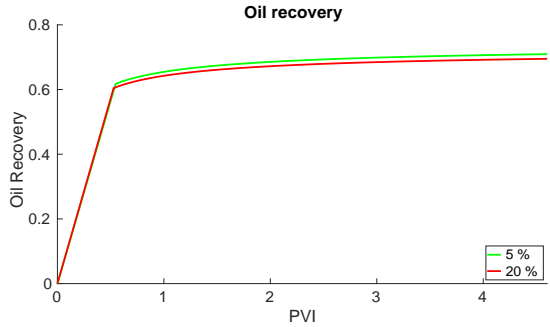}
	\caption{Evaluating oil recovery with a decrease of 5$\%$ and 20$\%$ in salt concentration.} 
	\label{fignova3}
\end{figure}

Figure \ref{fignova3a} displays the profiles of water saturation $S_w$, $pH$, and chloride $[Cl]$ concentration for the two examples. Figure \ref{fignova3} shows the corresponding oil recovery for a 5$\%$ and 20$\%$ reduction in the rock salt concentration. We observed that a 5$\%$ reduction in salt concentration resulted in an increase in the oil recovery fraction of 0.066 with respect to the case of a 20$\%$ of salt concentration reduction. The oil recovery values obtained with our model are consistent with those reported in the experiments, as shown in studies such as \cite{jerauld2008modeling} and \cite{korrani2019modeling}.

The Riemann solutions provide a numerical quantification and explanation of the influence of low salinity injection on the system state. In case (a) (as shown on the left side of Figure \ref{fignova3a}), we present the solution for a 5$\%$ reduction in the salt concentration, while in case (b) (as shown on the right side of Figure \ref{fignova3a}), we present the solution for a 20$\%$  reduction in the salt concentration. The Riemann solutions demonstrate the patterns of the system state, illustrating the effect of low salinity injection on the system.

In both cases (a) and (b) (as shown in Figure \ref{fignova3a}), we observe the formation of three fronts, i.e., saturation $S_w$, $pH$ and clhoride concentration $[Cl]$. The saline and $pH$ fronts move at the same velocity, while the saturation front has a slightly higher velocity. The velocity of the saline and $pH$ fronts is $4.05 \times 10^{-5}$ for both cases, whereas the velocity of the saturation front is $4.34 \times 10^{-5}$ for case (a) and $4.44 \times 10^{-5}$ for case (b).

The increase in $pH$ and the location of the front in $pH$ close to the saturation front can be explained by the fact that $pH$ controls the number of surface species at the interfaces of the oil/brine and the brine/carbonate (see e.g., \cite{xie2018ph}). This behavior of $pH$ and of the saline front can also be explained by the fact that $pH$ and salinity are among the most prominent factors affecting the wettability state of a crude oil/brine/rock system during waterflooding operations (\cite{mehraban2021experimental}).

Another experiment was conducted to calculate the residual oil by changing the types of oil 1 and 2 presented in Table \ref{tab1}. However, no significant changes were observed for the type of oil studied in this experiment.

We confirmed that the coupling of SCM and compositional modeling accurately reproduces the main effects observed in the experiments, including wettability, residual oil, and connate water saturation as the main factors responsible for oil recovery. The processes controlling wettability and surface complexes are the change in $pH$ and the formation of a $pH$ front.

\section{Conclusion}

This study quantifies the TBP of surface complexes and of wettability to estimate the effect of reducing rock salt concentration on permeability, and consequently, on oil recovery. We use a combination of SCM and compositional modeling to analyze the influence of wettability, residual oil, and connate water saturation on enhanced oil recovery. Our model reveals the presence of a shock in the saturation profile, with higher oil recovery being responsible for this phenomenon. We also observe a $pH$ wave in a constant $pH$ flood, salt formation at the front, and a jump in water saturation. The Riemann solution confirms the occurrence of $pH$ variations numerically using COMSOL and RPNfilho solvers, which is expected when surface complexes are formed.
 
\begin{sloppypar}

\section*{Acknowledgments}
The authors are gratefull to Ali A. Eftekhari for reviewing the calculations carried out using the PHREEQC program. Additionally, they would like to thank Sergio Pilotto for his support and acknowledge the funding received from CAPES under grant 88881.156518/2017-01 and CAPES/NUFFIC grant 88887.156517/2017-00, CNPq under grants 405366/2021-3 and 306566/2019-2, and FAPERJ under grants E-26/210.738/2014, E-26/202.764/2017, and E-26/201.159/2021.

\end{sloppypar}

\section{Declarations}
There are no conflicts of competing interests.

\bibliographystyle{unsrt}

\end{document}